\documentstyle[12pt]{article}

\renewcommand{\thefootnote}{\fnsymbol{footnote}}

\catcode`\@=11
\def\maketitle{\par
 \begingroup
 \def\thefootnote{\fnsymbol{footnote}}
 \def\@makefnmark{\hbox
 to 0pt{$^{\@thefnmark}$\hss}}
 \if@twocolumn
 \twocolumn[\@maketitle]
 \else \newpage
 \global\@topnum\z@ \@maketitle \fi\thispagestyle{empty}\@thanks
 \endgroup
 \setcounter{footnote}{0}
 \let\maketitle\relax
 \let\@maketitle\relax
 \gdef\@thanks{}\gdef\@author{}\gdef\@title{}\let\thanks\relax}
\def\@maketitle{\newpage
 \null
 \hbox to\textwidth{\hfil\hbox{\begin{tabular}{r}\@preprint\end{tabular}}}
 \vskip 2em \begin{center}
 {\Large\bf \@title \par} \vskip 1.5em {\normalsize \lineskip .5em
\begin{tabular}[t]{c}\@author
 \end{tabular}\par}
 \end{center}
 \par
 \vskip 1.5em}
\def\preprint#1{\gdef\@preprint{#1}}
\def\abstract{\if@twocolumn
\section*{Abstract}
\else \normalsize
\begin{center}
{\large\bf Abstract\vspace{-.5em}\vspace{0pt}}
\end{center}
\quotation
\fi}
\def\endabstract{\if@twocolumn\else\endquotation\fi}
\catcode`\@=12

\begin{document}
\baselineskip=.285in

\preprint{SNUTP-97-088\\[-2mm] gr-qc/9707056}

\title{\Large\bf Global Vortex and Black Cosmic String
\protect\\[1mm]\ }
\author{{\normalsize Nakwoo Kim, Yoonbai Kim$^{\ast}$ and Kyoungtae Kimm}\\
{\normalsize\it Department of Physics and Center for Theoretical Physics,}\\
{\normalsize\it Seoul National University, Seoul 151-742, Korea}\\
{\normalsize\it nakwoo$@$phya.snu.ac.kr, dragon$@$phya.snu.ac.kr}\\
{\normalsize\it $^{\ast}$Department of Physics, Sung Kyun Kwan University, 
Suwon 440-746, Korea}\\
{\normalsize\it yoonbai$@$cosmos.skku.ac.kr}}
\date{}
\maketitle

\renewcommand{\theequation}{\thesection.\arabic{equation}}

\begin{center}
{\large\bf Abstract}\\[3mm]
\end{center}
\begin{quote}
\hspace{7mm} We study global vortices coupled to (2+1) dimensional 
gravity with negative cosmological constant. 
We found nonsingular vortex solutions in $\phi^4$-theory with a broken 
$U(1)$ symmetry, of which the spacetimes do not
involve physical curvature singularity. When the magnitude of negative 
cosmological constant is larger than a critical value at a given symmetry
breaking scale, the spacetime structure is a regular hyperbola, however
it becomes a charged black hole when the magnitude of cosmological constant
is less than the critical value.
We explain through duality transformation the reason why
static global vortex which is electrically neutral forms 
black hole with electric charge.
Under the present experimental bound of the cosmological constant,
implications on cosmology as a straight black cosmic string is also discussed
in comparison with global $U(1)$ cosmic string in the spacetime of
the zero cosmological constant.

\noindent PACS number(s): 11.27.+d, 04.40-b, 04.70.Bw
\end{quote}
\newpage
\pagenumbering{arabic}
\thispagestyle{plain}
\setcounter{section}{1}
\begin{center}\section*{\large\bf I. Introduction}\end{center}
\indent\indent 
Einstein gravity in (2+1) dimensions has no local degrees of freedom and the 
matter coupled to gravity changes only the global structure of spacetime 
outside sources~\cite{DJH}. 
Subsequently anti-de Sitter solutions in three dimensional 
gravity were analyzed in eighties~\cite{DJ}, 
however it took many years thenceforth to find out the black hole 
structure among those solutions of negative 
cosmological constant~\cite{BTZ}.

One of the reasons why conic solutions formed by point particles in (2+1)D 
have attracted attention is that they stand for the asymptotic space of 
cylindrically symmetric local cosmic strings 
which are extended solitonic objects~\cite{VS}. 
In this context, an intriguing question in (2+1) dimensions
with the negative cosmological constant is 
whether one can find the structure of black cosmic strings or not.
In relation with the stability of such string-like objects in (3+1)D,
topological vortex solution in (2+1)D is the first candidate.
There has been another subject in (3+1) dimensions:
The study of black holes, particularly the charged black holes,
formed by the solitons, {\em e.g.}, monopoles, Skyrmions, etc,
has been an interesting subject~\cite{LNW}.
Since the negative cosmological constant amounts to the term of energy
proportional to the area of spatial manifold ($\sim r^2$), one can
easily guess that static extended objects carrying long-range tail are
important.
The simplest candidate in (2+1) dimensions may 
be the global $U(1)$  vortex
of which energy diverges logarithmically in flat spacetime,
which is a viable cosmic string candidate 
in cosmology~\cite{VS,Gre,HS}.
Here if we remind of a fact that a physical curvature singularity in
global string spacetime is unavoidable in case of the zero cosmological 
constant, 
we can add another question whether
we can find regular global cosmic strings in anti-de Sitter spacetime or 
black cosmic strings with no divergent curvature~\cite{KKK}.

In this paper, we will consider a complex scalar $\phi^4$ model in
(2+1) dimensional anti-de Sitter spacetime and look for the global
vortex solutions. There are cylindrically symmetric global $U(1)$
vortex solutions connecting smoothly the symmetric local
maximum at the origin and a broken vacuum point at 
spatial infinity. 
The spacetimes formed by these strings are regular hyperbola
with deficit angle, extremal black hole and charged black hole
as the magnitude of the cosmological constant decreases.
The curvature of these solutions is not divergent everywhere even for the
charged black holes. 
So it is contrary to the zero cosmological constant case, 
where the $U(1)$ global string admits no globally well-behaved
solution.

This paper is organized as follows.
We begin in Sec. II by establishing explicitly the relation 
that the (2+1)D spinless black hole solutions in Ref.~\cite{BTZ}
are part of general anti-de Sitter space solutions in Ref.~\cite{DJ}.
In Sec. III, we introduce the model and obtain the global $U(1)$
vortex solutions. 
Possible geodesics of massless and massive test particles are also given.
In Sec. IV, the connection between the topological charge and the electric 
charge of black hole is illustrated by use of the duality transformations.
In Sec. V, questions on the physical relevance as a charged black cosmic string
in (3+1) dimensions are addressed. 
We conclude in Sec. VI with a brief discussion.
\setcounter{section}{2}
\setcounter{equation}{0}
\begin{center}
\section*{\large\bf II. Black Hole as an Anti-de Sitter Solution}
\end{center}
\indent\indent 
In this section, let us recapitulate what the (2+1) dimensional Schwarzschild 
black hole solution is among a series of anti-de Sitter solutions of which all
static metrics can be characterized in terms of one complex function.
Under conformal gauge, the static metric compatible with static objects is
parameterized by
\begin{eqnarray}\label{metr}
ds^{2}=\Phi^{2}(z,\bar z)dt^{2}-b(z,\bar z)dzd\bar{z},
\end{eqnarray}
where $z\equiv x+iy=Re^{i\Theta}$. 
For $n$ massive spinless point particles 
located at positions $z=z_{a},\;a=1,2,\cdots,n$, each
with mass $m_{a}$, 
the cosmological constant $\Lambda$ is obtained by solving 
the time-time component of 
Einstein equations\footnote{Our equation in Eq.~(\ref{eq00}), 
which uses the action for the
point particles
\begin{eqnarray}
S_{pp}=\sum^{n}_{a=1}m_{a}
\int ds\sqrt{g^{\mu\nu}(x_{a}(s))\frac{dx_{\mu}^{a}}{ds}
\frac{dx_{\nu}^{a}}{ds}}\nonumber ,
\end{eqnarray}
is different from the equation written Eq.~(2.7) in Ref.\cite{DJ}.
Specifically $m_{a}=\Phi(x)^{2}m_{a}^{DJ}$. For the point sources, these
two equations lead to the same solutions which were firstly obtained by the
authors in Ref.\cite{DJ}. However there is a possibility that the solutions 
for the extended sources may include different curved spacetime. Y.K. would 
like to thank R. Jackiw for the discussion on this point.} 
\begin{eqnarray}\label{eq00}
\Lambda=-\frac{2}{b}\partial_{z}\partial_{\bar{z}}
\ln b\sum_{a=1}^{n}|z_{a}-z|^{8Gm_{a}}.
\end{eqnarray}
The space-space components give two independent equations: 
One is for the spatial trace~\cite{KK}
\begin{eqnarray}\label{ijtr}
\Lambda=-\frac{2}{\Phi b}\partial_{z}\partial_{\bar{z}} \Phi,
\end{eqnarray}
and the other is for traceless part
\begin{eqnarray}\label{ijle}
\partial_z\Bigl(\frac{1}{b}\partial_z \Phi\Bigr)=0.
\end{eqnarray}
As obtained in Ref.\cite{DJ},
the general anti-de Sitter solution of Eq.~(\ref{eq00}), Eq.~(\ref{ijtr}) and 
Eq.~(\ref{ijle}) is
\begin{eqnarray}\label{beq0}
b&=&\frac{\varepsilon}{|\Lambda|V(z)\bar{V}(\bar{z})\sinh^{2}\sqrt{\varepsilon}
(\zeta-\zeta_{0})}\\
\label{phieq0}
\Phi&=&\sqrt{\varepsilon}\coth\sqrt{\varepsilon}(\zeta-\zeta_{0}),
\end{eqnarray}
where $V(z)$ ($\bar{V}(\bar{z})$) is an 
arbitrary (anti-)holomorphic function 
and  $\zeta$ is a real variable defined by
\begin{eqnarray}
\zeta\equiv\frac{1}{2}\biggl[\int^{z}\frac{dw}{V(w)}+\int^{\bar{z}}
\frac{d\bar{w}}{\bar{V}(\bar{w})}\biggr].
\end{eqnarray}
$\varepsilon$ is a real positive integration constant for $\Lambda>0$ and 
is an arbitrary nonzero real constant for $\Lambda<0$. 

Here let us consider the simplest case that $V=z/c$ where $c$ is a real
constant so as to keep the single-valuedness of $\zeta$.
When $\varepsilon>0$, one can set $\varepsilon=1$ without loss of any
generality. In the radial coordinate, $c$ is identified as $c=1-4Gm$
and the metric in 
Eq.~(\ref{metr}) becomes
\begin{eqnarray}\label{demet}
ds^{2}=\biggl(\frac{R^{(1-4Gm)}+R^{-(1-4Gm)}}{R^{(1-4Gm)}-R^{-(1-4Gm)}}
\biggr)^{2}dt^{2}-\frac{4(1-4Gm)^{2}}{|\Lambda|R^{2}(R^{(1-4Gm)}
-R^{-(1-4Gm)})^{2}}(dR^{2}+R^{2}d\Theta^{2}), 
\end{eqnarray}
where $m$ is the total mass of the point particle at $R=0$. Introducing the new
coordinates $r$ and $\theta$ such as
\begin{eqnarray}
r=\frac{2}{|\Lambda|^{1/2}}\frac{1}{|R^{(1-4Gm)}-R^{-(1-4Gm)}|} 
~~\mbox{and}~~\theta=(1-4Gm)\Theta,
\end{eqnarray}
we can rewrite the metric in Eq.~(\ref{demet}) as
\begin{eqnarray}\label{demet2}
ds^{2}=(1+|\Lambda|r^{2})dt^{2}-(1+|\Lambda|r^{2})^{-1}dr^{2}-r^{2}d\theta^{2}.
\end{eqnarray}
Now we can easily identify the structure of manifold as a hyperbola with
deficit angle $\delta=8\pi Gm$ where $4Gm<1$. 

The above is the physical
interpretation provided in Ref.\cite{DJ}. Then, how about the solutions
with negative $\varepsilon$? 
Or equivalently, the $\varepsilon=-1$ case? In this case, the metric
in Eq.~(\ref{metr}) can be reexpressed as
\begin{eqnarray}\label{demet3}
ds^{2}=\frac{1}{\tan^{2}(2c\ln R)}dt^{2}-\frac{c^{2}}{|\Lambda|\sin^{2}
(2c\ln R)}(d\ln R^{2}+d\theta^{2}).
\end{eqnarray}
One can easily notice that the coordinate $R$ has unconventional ranges such
that $\Phi(R)$ and $R^{2}b(R)$ diverge at $R=\exp\bigl(\frac{k\pi}{4c}\bigr)$
($k$ is an integer). Because of the unconventional behavior of metric
functions at $R=0$, it seems rather difficult to pin the unknown constant
$c$ down by use of the point particle mass $m$ under this coordinate system.
To make physics clear, let us do a coordinate transformation:
\begin{eqnarray}
r=\frac{c}{|\Lambda|^{1/2}\sin(2 c\ln{R})}.
\end{eqnarray}
The result leads to the well-known 
(2+1)D (Schwarzschild type) black hole solution
with mass $c^{2}$ and negative cosmological constant $\Lambda$ \cite{BTZ}:
\begin{eqnarray}\label{Schw}
ds^{2}=(-c^2+|\Lambda|r^2)dt^{2}
-(-c^2+|\Lambda|r^2)^{-1}dr^{2}-R^2d\theta^{2}.
\end{eqnarray} 
As shown in Fig.~1, each range of $r$
$\Big(\exp\frac{k\pi}{4c}<r<\exp\frac{(k+1)\pi}{4c}\Big)$ 
covers the exterior region of
the Ba\~nados-Teitelboim-Zanelli(BTZ) solution.

\begin{figure}

\setlength{\unitlength}{0.1bp}
\begin{picture}(3600,2160)(0,0)
\put(1900,20){\makebox(0,0){{\Large$c\ln R$}}}
\put(100,1180){%
\makebox(0,0)[b]{{{\Large$r$}}}%
}
\put(3437,200){\makebox(0,0){$2\pi$}}
\put(2668,200){\makebox(0,0){$3\pi/2$}}
\put(1900,200){\makebox(0,0){$\pi$}}
\put(1131,200){\makebox(0,0){$\pi/2$}}
\put(363,200){\makebox(0,0){0}}
\put(283,320){\makebox(0,0){0}}
\put(213,900){\makebox(0,0){$|c\Lambda|^{\frac{1}{2}}$}}
\end{picture}

\label{fig1}
\caption{The radial coordinate $R$ under the conformal gauge vs. 
$r$ of the Schwarzschild metric.}
\end{figure}
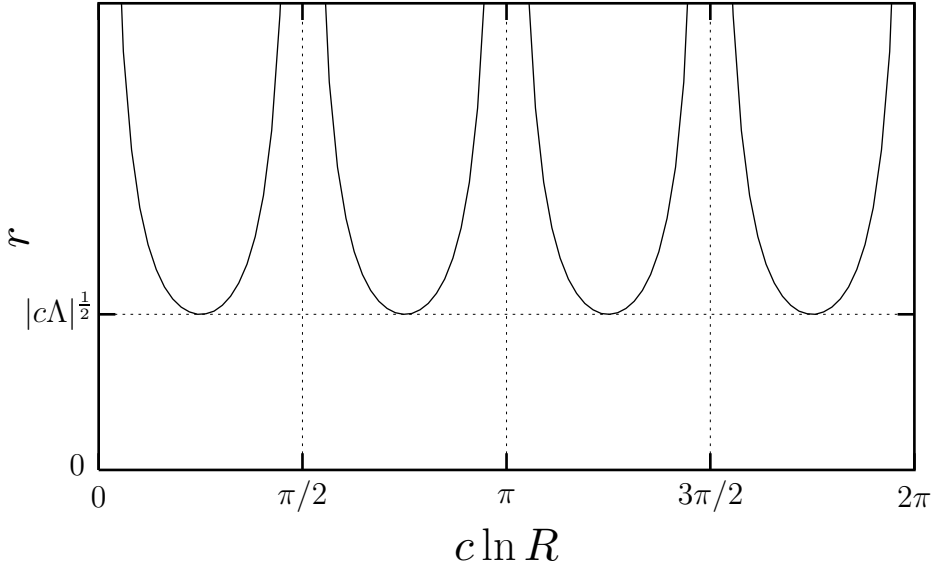

One step extension may be the quadratic $V(z)$, i.e.,
$V=(z-z_{1})(z-z_{2})/\tilde{c},\;(z_{1}\neq z_{2})$. 
Reexamining the above computation, we can
easily conclude that this reproduces the same result again; (i) when
$\varepsilon=1$, it goes to anti-de Sitter space with point particle mass 
$4Gm=1-\Bigl|\frac{\tilde{c}}{z_{1}-z_{2}}\Bigr|$, and
(ii) when $\varepsilon=-1$, it goes to the BTZ black hole with the black hole 
mass $8GM=\Bigl|\frac{\tilde{c}}{z_{1}-z_{2}}\Bigr|^{2}$.
For more complicated examples, further study is needed. 

It is now the turn of charged case.
The metric compatible with such spinless static objects and with rotational
symmetry is of a form
\begin{eqnarray}\label{metric}
ds^2=e^{2N(r)} B(r) dt^2-\frac{1}{B(r)}dr^2 -r^2d\theta^2.
\end{eqnarray}
Then Einstein equations are
\begin{eqnarray}
\label{neq}
\frac{1}{r}\frac{dN}{dr}&=& \frac{8\pi G}{B}(T^t_{\;t}-T^r_{\;r})\\
\label{beq}
\frac{1}{r}\frac{dB}{dr}&=&2|\Lambda|-16\pi GT^t_{\;t}.
\end{eqnarray}

If the source is composed of the electrostatic field of a point charge $q$, the 
energy-momentum density is
\begin{eqnarray}\label{elec}
T^{t}_{\;t}=T^{r}_{\;r}=\frac{e^{-2N(r)}}{2}E_{i}^{2}
=\frac{q^{2}}{2}\frac{e^{-2N(r)}}{r^{2}}.
\end{eqnarray}
Inserting Eq.~(\ref{elec}) into the Einstein equations in Eq.~(\ref{beq}) and 
Eq.~(\ref{neq}), we have 
\begin{eqnarray}
N(r)&=&0\\
B(r)&=&|\Lambda|r^{2}-8\pi Gq^{2}\ln r -8GM,
\label{RNbh}
\end{eqnarray}
where $M$ is an undetermined mass parameter (of which dimension is 
mass per unit length in (3+1)D).
When $M\geq \frac{\pi}{2} q^{2}(1-\ln \frac{4\pi Gq^{2}}{|\Lambda|})$,
the Reissner-Nordstr\"om type black hole 
with two horizons are formed, and the 
extremal one is formed  when the equality holds. For the
Schwarzschild type black hole case, there is no 
curvature singularity at any $r\;(r\geq 0)$. 
However, this charged black hole contains the curvature singularity at
the origin, $R^{t}_{\;t}=6|\Lambda|-8\pi G q^2/r^2$ 
due to the infinite self energy of the point charge.
It should be noted that the black hole charge is generated by the so-called 
logarithmically divergent energy term, 
$\int_{\epsilon}^{L} drrT^{t}_{\;t}\sim \ln L$,
at large distance
but the ultraviolet singularity at the origin ($\epsilon\rightarrow 0$) 
is not essential for the  
formation of the charged black hole even though 
it gives a divergent curvature. 
Therefore, the global vortex attracts our interest since it
involves the long-tail of energy density despite it tames
the ultraviolet divergence at the vortex core of which 
the singularity is irrelevant to the black hole structures.

We have an explicit solution of charged BTZ black hole, so the question is
whether there are two or more solutions under the conformal gauge in 
Eq.~(\ref{metr}), of which one corresponds to the charged BTZ solution. Again,
let us consider the spacetime geometry when a point particle of mass $m$ 
and charge $q$ sitting at the origin. 
The relevant Einstein equation is:
\begin{eqnarray}\label{emeq}
\Lambda= -\frac{2}{b}\partial_z\partial_{\bar z}\ln{b}
|z|^{8\pi G m}-16\pi G T^{t}_{em\;t} ,
\end{eqnarray}
and the energy density distribution determined by the Gauss' law, namely
$T^{t}_{em\;t}=\frac{1}{2b}(q^2/R^2)$. Again Eq.~(\ref{emeq}) 
with negative cosmological constant reduces to a
Liouville type equation:
\begin{eqnarray}
2\partial_{z}\partial_{\bar{z}}\ln\Bigl(\frac{b}{ R^{-8Gm-4\pi Gq^2\ln{R}}}  
\Bigr)=|\Lambda| b.
\end{eqnarray}
It is known that this equation is not integrable and there is no
known exact solution of this equation, yet~\cite{Inc}. 
Once we try to do a coordinate
transformation from Eq.~(\ref{metr}) to Eq.~(\ref{elec}) and Eq.~(\ref{RNbh}),
the reason why we can not obtain the explicit form of the solution in the
conformal metric is obvious 
though we could get it in the Schwarzschild metric. 
Despite the algebraic relation $R^2b(R)=r^2$, $R$ can not be
expressed as a function of $r$ in a closed form for the charged objects:
\begin{eqnarray}
r^{2}=r^{2}_{0}\exp\bigg\{\int^{r}
\frac{d\ln R^{2}}{\sqrt{|\Lambda|R^{2}-4\pi Gq^{2}\ln R^{2}-8GM}}\bigg\}.
\end{eqnarray}
Only when the object is neutral ($q=0$), this integral is done
in a closed form. 
{}For a negative $M$, the integration range of $r$ is not restricted
and then we have the hyperbolic solution in Eq.~(\ref{demet}),
namely a positive $\epsilon$ solution in Eq.~(\ref{beq0}).
On the other hand, for a positive $M$, the integration range of $r$
larger than $M/|\Lambda|$ and it is the BTZ black hole solution outside
horizon in Eq.~(\ref{Schw}), namely a negative $\epsilon$ solution.
Though we do not know the closed form of the metric function in conformal
coordinate and do not determine the range of integration range of $r$
explicitly, it is obvious that there are both types of solutions for charged
case ($q\ne 0$).
When $M<\frac{\pi}{2}q^2(1-\ln\frac{4\pi Gq^2}{|\Lambda|})$, we have a
regular solution corresponding to the positive $\epsilon$ solution. 
When $M>\frac{\pi}{2}q^2(1-\ln\frac{4\pi Gq^2}{|\Lambda|})$,
the obtained metric describes the inside of inner horizon and outside of 
outer horizon of the charged black hole corresponding to the negative
$\varepsilon$.
Note that the positivity of $M$ is not a necessity for the charged 
black hole ($q\ne 0$).
We will show it is indeed the case in global $U(1)$ vortices.

In this section we have clarified a relation between the rotationally
symmetric solutions in the metric under conformal gauge and the static BTZ
black hole solutions in the Schwarzschild type metric. However, the similar
construction like the above relation is not
clear for more complicated solutions, {\it e.g.}, 
multicenter solutions without 
rotational symmetry, the solutions with the deficit angle 
equal to or bigger than $2\pi$, 
and the spinning black holes~\cite{Cle}, yet.  

\setcounter{section}{3}
\setcounter{equation}{0}
\begin{center}\section*{\large\bf III. Global Vortex in Anti-de Sitter Space as a 
Regular Neutral Vortex or a Charged Black Hole}
\end{center}
\indent\indent 
Let us consider the anti-de Sitter spacetime in the presence of global
vortices. 
The standard example is given by action:
\begin{eqnarray}\label{action}
S=\int d^{3}x\sqrt{g}\,\biggl\{-\frac{1}{16\pi G}(R+2\Lambda)+
\frac{1}{2}g^{\mu\nu}\partial_{\mu}\bar{\phi}\partial_{\nu}\phi
-\frac{\lambda}{4}(\bar{\phi}\phi-v^{2})^{2}\biggr\},
\end{eqnarray}
where $\phi(x)$ is a complex field.
The ansatz for the static global vortices with rotational symmetry is
\begin{eqnarray}\label{ansatz}
\phi=|\phi|(r)e^{in\theta}.
\end{eqnarray}
{}From the model given above in Eq.~(\ref{action}), the equation for
scalar field is
\begin{eqnarray}\label{feq}
\frac{d^{2}|\phi|}{dr^{2}}+\Bigl(\frac{dN}{dr}+\frac{1}{B}\frac{dB}{dr}
+\frac{1}{r}\Bigr)\frac{d|\phi|}{dr}&=&
\frac{1}{B}\Bigl(\frac{n^2|\phi|}{r^2} 
+\lambda (|\phi|^2-v^2)|\phi|\Bigr).
\end{eqnarray}
The energy-momentum tensor for the Einstein equations 
includes the long tail term ($\sim 1/r^2$):
\begin{eqnarray}
T^{t}_{\;t}&=&\frac{1}{2}\biggl\{B\biggl(\frac{d|\phi|}{dr}\biggr)^2
+\frac{n^2}{r^2}|\phi|^{2}
+\frac{\lambda}{2}(|\phi|^2-v^2)^2\biggr\}
\label{ttt}\\
T^r_{\;r}&=&\frac{1}{2}\biggl\{-B\biggl(\frac{d|\phi|}{dr}\biggr)^2
+\frac{n^2}{r^2}|\phi|^{2}
+\frac{\lambda}{2}(|\phi|^2-v^2)^2\biggr\}.
\label{trr}
\end{eqnarray}
Substituting Eq.~(\ref{ttt}) and Eq.~(\ref{trr})
into the Einstein equations in Eq.~(\ref{neq}) and Eq.~(\ref{beq}), 
we obtain the following equations:
\begin{eqnarray}
\label{Neq}
\frac{1}{r}\frac{d N}{d r} &=&8\pi G \Big( \frac{d |\phi|}{dr}\Big)^2 \\
\frac{1}{r}\frac{dB}{dr}&=&2|\Lambda| 
- 8\pi G\biggl\{ B\Bigl(\frac{d|\phi|}{dr}\Bigl)^{2}+\frac{n^2}{r^2}|\phi|^2
 +\frac{\lambda}{2}(|\phi|^2-v^2)^2\biggr\},
\label{Beq}
\end{eqnarray}
and then the metric functions $N(r)$ and $B(r)$ 
are expressed in terms of the scalar field:
\begin{eqnarray}\label{nreq}
N(r)= -8\pi G\int^{\infty}_{r}dr'r'\Big( \frac{d |\phi|}{dr'}\Big)^2
\end{eqnarray}
\begin{eqnarray}\label{breq} 
B(r) &=& e^{-N(r)} \Bigg\{ 2|\Lambda|\int^{r}_{0}  dr' r' 
\exp\Big[-8\pi G \int^\infty_{r'} dr''
        r''\Big(\frac{d|\phi|}{dr''}\Big)^2\Big] \\
    \nonumber 
  &&\hspace{0mm} -8\pi G\int^r_0 dr'r'
\exp\Big[-8\pi G \int^\infty_{r'} dr'' r''
\Big(\frac{d|\phi|}{dr''}\Big)^2\Big]
\Big[\frac{n^2}{r'^2}|\phi|^2
 +\frac{\lambda}{2}(|\phi|^2-v^2)^2\Big]
+  e^{N(0)}\Bigg\}.
\end{eqnarray}

Here we choose a set of boundary conditions, $B(0)=1$ and $N(\infty)=0$, 
according to the following reason: When we take the limit of both no matter
$(T^{\mu}_{\;\nu}=0)$ and zero vacuum energy $(\Lambda=0)$, the
spacetime reproduces Minkowski spacetime. Since a rescaling of radial
coordinate $r$ leads to a flat cone with deficit angle $2\pi(1-\sqrt{B(0)}\;)$
in this limit,
$B(0)=1$ is an appropriate choice. If the coincidence of propertime for the
observer at spatial infinity is asked, then the temporal coordinate $t$ selects
$N(\infty)=0$. 
In the context of scalar field, the configuration of our interest is the
solitonic one approaching its vacuum value at spatial infinity, i.e.,
$|\phi|(\infty)=v$. Now one remaining boundary condition is about the scalar
amplitude at the origin. If there is no coordinate and curvature singularity, 
single-valuedness of scalar field forces 
$|\phi|(0)=0$ for the vortex solution ($n\ne 0$).
However, when we take into account the geometry with curvature
singularity or a black hole including
the horizons, it is not necessary in general for 
the scalar field configuration to be nonsingular. 
Such singular solutions, so called exotic black holes, 
that their scalar fields do no vanish at the origin have been studied in 
(3+1)D~\cite{LNW}.
However, it seems that there is a difference between the (3+1)D
black holes and the BTZ solution: 
The mass accumulated at the core of the black hole induces the steep curvature
change around its core and is crucial to make 
black hole in (3+1)D curved spacetime,
but the Schwarzschild-type BTZ solution does
not have any signal of such accumulation singularity and the divergent 
curvature at the origin of
Reissner-Norstr\"om-type BTZ solution in Eq.~({\ref{RNbh}}) 
is irrelevant to the black hole
structure as explained previously. 
In this respect, an intriguing  question is 
whether there is the global vortex solution interpolating 
smoothly $|\phi|(0)=0$ and $|\phi|(\infty)=v$ even  
in asymptotically anti-de Sitter spacetime.
As mentioned previously, 
the charged BTZ black hole made by the electric point charge involves an
unnecessary divergent curvature at the origin, the regular extended
objects, specifically the neutral global vortex, can have a chance
to form a curvature-singularity-free charged BTZ black hole.

The question whether there exist smooth vortex configurations
or not is also intriguing in the context of 
no-go theorem that this global $U(1)$ scalar model can not
support finite energy static regular vortex configuration in flat spacetime.
Thus the global $U(1)$ vortex carries logarithmically divergent energy.
This symptom seems to appear in curved spacetime that
the global $U(1)$ vortex does not admit globally well-behaved solution
when $\Lambda=0$~\cite{Gre}. The negative potential energy (the
negative cosmological constant) comes in this model through the coupling of 
gravity although it does not have its own propagating degrees in
(2+1)D. Therefore, one may expect the existence of regular vortex
configurations in anti-de Sitter spacetime, and we will show that it is indeed the
case in the global $U(1)$ model of our interest.

Near the origin, the power series solutions up to the leading term are
\begin{eqnarray}
|\phi|(r)&\sim&\phi_{0}r^{n}
\label{phi00}\\
N(r)&\sim& N(0)+4\pi G n \phi_0^2 r^{2n}\\
B(r)&\sim&1+(|\Lambda| -2\pi G \lambda v^4 -8\pi G \phi_0^2 \delta_{n1})r^2 .
\label{br0}
\end{eqnarray}
Since the right-hand side of Eq.~(\ref{Neq}) is positive definite,
$N(r)$ is monotonically increasing everywhere.
Eq.~(\ref{br0}) tells us that, when the scale of cosmological constant
$|\Lambda|/\lambda v^2$ is smaller that the Planck scale $2\pi G v^2$,
then $B(r)$ starts to decrease near the origin.

Though we will take into account the geometry with the
``horizon" and it may hinder the systematic expansion of the solution, let us
attempt the power series solution up to the leading term for sufficiently 
large $r$:
\begin{eqnarray}
|\phi|(r)&\sim&v-\frac{\phi_{\infty}}{r^{2}}\\
N(r)&\sim&-\frac{8\pi G\phi^{2}_{\infty}}{r^{4}}
\label{ninf}\\
B(r)&\sim&|\Lambda|r^{2}-8\pi Gv^{2}n^{2}\ln{r/r_{c}}- 8G{\cal M}+1
+{\cal O}(1/r^{2}),\label{binf}
\end{eqnarray}
where ${\cal M}$ is the integration constant.
Let us estimate the core mass ${\cal M}$ and the 
core size $r_{c}$ of the global vortex. 
As a simple but valid approximation, let us assume
\begin{eqnarray}\label{app}
|\phi|(r)=\left\{\begin{array}{ll}
0 & \mbox{when} ~~r< r_{c}\\
v & \mbox{when} ~~r\geq r_{c}
\end{array}\right.
\end{eqnarray}
and neglect the change of the metric function $N(r)$, i.e., $N\sim 0$.
Substituting Eq.~(\ref{app}) into Eq.~(\ref{breq}),
we have 
\begin{eqnarray}\label{mcore}
B(r)-|\Lambda|r^2-8\pi Gv^2 n^2\ln{r} -1 &\approx&
2\pi G v^2(\lambda v^2 r^2_{c} -4n^2\ln{r_{c}}) \nonumber \\
&\geq& 4\pi G v^2 n^2 (1-\ln{2n^2/\lambda v^2}) .
\end{eqnarray}
Here the minimum value in the second line of Eq.~(\ref{mcore}) 
is obtained when 
$r_{c}\sim n/\sqrt{\lambda} v$ and ${\cal M}\sim \frac{\pi}{2} v^2 n$.
Crude as this approximation is, one can read the minimum point of $B(r)$:
\begin{eqnarray}
r_m=\sqrt{\frac{4\pi G v^2n^2}{|\Lambda|}}
\label{roughrm}
\end{eqnarray}
which may be valid when the minimum position $r_m$ is much larger than the 
core radius.
The positivity of core mass $\cal M$ after subtracting the logarithmic long tail
is different from the global monopole in (3+1)D
curved spacetime with zero cosmological constant.
The core mass of global monopole is negative and
the repulsive nature at the monopole core leads to the
impossibility of the formation of global monopole-black hole
even at the Plank scale~\cite{HL}.

Suppose that there exists a horizon, namely, the position $r_{H}$
where the metric function $B(r)$ vanishes.
At the horizon, the boundary conditions are from 
Eq.~(\ref{feq}) and Eq.~(\ref{Beq}):
\begin{eqnarray}
\left\{\begin{array}{ll}
(i) & B_{H}=0 \\
(ii) & {\displaystyle
\frac{d|\phi|}{dr}\Bigg|_{H}=
\frac{|\phi|_H\Big[\frac{n^2}{r_H^2} +\lambda(|\phi|^2_H-v^2)\Big]}{
8\pi Gr_H\Big[\frac{|\Lambda|}{4\pi G}-\Big( \frac{n^2}{r_H^2}|\phi|^2_H 
+\frac{\lambda}{2}(|\phi|^2_H-v^2)^2\Big)\Big]}     },
\end{array}\right.
\label{dfdr}
\end{eqnarray}
where $B_H=B(r_H)$ and $|\phi|_H =|\phi|(r_H)$.
The behaviors of functions near the horizon are approximated by
\begin{eqnarray}
|\phi|(r)&\sim&|\phi|_H+\frac{d|\phi|}{dr}\Bigg|_H(r-r_{H}) +
\frac{1}{2} \frac{d^2 |\phi|}{dr^2}\Bigg|_H (r-r_H)^2\\
N(r)&\sim& N_H +N_{H1} (r-r_H) + N_{H2} (r-r_H)^2 \label{nrh}\\
B(r)&\sim&B_{H1}(r-r_{H})+B_{H2}(r-r_{H})^{2},
\label{brh}
\end{eqnarray}
where $N_H=N(r_H)$,  
$\frac{d^2|\phi|}{dr^2}\Big|_H$, $N_{H1}$
$N_{H2}$, $B_{H1}$ and $B_{H2}$ are expressed in terms of 
$|\phi|_H$ at $r_{H}$:
\begin{eqnarray}
\frac{d^2|\phi|}{dr^2}\bigg|_H&=&
\frac{|\phi|_H\Big[\frac{n^2}{r_H^2}+\lambda(|\phi|_H^2 -v^2)^2\Big]}
{64\pi^2G^2 r_H^2\Big[\frac{|\Lambda|}{4\pi G}
-\Big(\frac{n^2}{r_H^2}|\phi|_H^2 
+\frac{\lambda}{2}(|\phi|_H^2-v^2)^2\Big)\Big]^2}
\Bigg\{-8\pi G\bigg\{ \frac{n^2}{r^2_H} |\phi|^2_H +
\nonumber \\
&& \bigg[
\frac{|\Lambda|}{4\pi G}
-\Big(\frac{n^2}{r_H^2}|\phi|^2_H +\frac{\lambda}{2}(|\phi|_H^2 -v^2)^2
\Big)\bigg]\bigg[1
+\frac{n^2}{r_H^2}\frac{1}{\Big[\frac{n^2}{r_H^2}
                     +\lambda (|\phi|^2_H -v^2)^2\Big]}
\bigg]\bigg\}
\nonumber \\
&& +\frac{3|\phi|_H^2\Big[ \frac{n^2}{r_H^2}
     + \lambda(|\phi|_H^2 -v^2)^2\Big]^2}{
 2\Big[ \frac{|\Lambda|}{4\pi G} -\Big( \frac{n^2}{r_H^2} |\phi|^2_H
 +\frac{\lambda}{2}(|\phi|_H^2 -v^2)^2\Big)\Big]}
+\frac{1}{2}\Big[ \frac{n^2}{r_H^2}
 +\lambda (3 |\phi|_H^2 -v^2)\Big]\Bigg\} \\
N_{H1}&=&
\frac{1}{8\pi G r_H}
\frac{ |\phi|_H^2 \Big[ \frac{n^2}{r_H^2}+\lambda(|\phi|_H^2-v^2)\Big]^2}
{\Big[\frac{|\Lambda|}{4\pi G} -\Big(\frac{n^2}{r_H^2}|\phi|_H^2 
+\frac{\lambda}{2}(|\phi|_H^2-v^2)^2\Big)\Big]^2 }\\
N_{H2}&=&
\frac{|\phi|^2_H \Big[ \frac{n^2}{r^2_H} + \lambda (|\phi|^2_H 
-v^2)^2 \Big]^2}{64 \pi^2 G^2 r^2_H \Big[
\frac{|\Lambda|}{4\pi G} - \Big( \frac{n^2}{r^2_H} |\phi|^2_H
+\frac{\lambda}{2} (|\phi|^2_H-v^2 \Big)^2 
\Big]^3 } 
\Bigg\{ -8\pi G \bigg\{ \frac{n^2}{r^2_H} |\phi|^2_H +
\nonumber \\
&& \bigg[ 
\frac{|\Lambda|}{4\pi G} 
-\Big(\frac{n^2}{r_H^2}|\phi|^2_H +\frac{\lambda}{2}(|\phi|_H^2 -v^2)^2
\Big)\bigg]\bigg[\frac{1}{2}
+\frac{n^2}{r_H^2}\frac{1}{\Big[\frac{n^2}{r_H^2}
                     +\lambda (|\phi|^2_H -v^2)^2\Big]}
\bigg]\bigg\}
\nonumber \\
&& +\frac{3|\phi|_H^2\Big[ \frac{n^2}{r_H^2}
     + \lambda(|\phi|_H^2 -v^2)^2\Big]^2}{
 2\Big[ \frac{|\Lambda|}{4\pi G} -\Big( \frac{n^2}{r_H^2} |\phi|^2_H
 +\frac{\lambda}{2}(|\phi|_H^2 -v^2)^2\Big)\Big]}
+\frac{1}{2}\Big[ \frac{n^2}{r_H^2} 
 +\lambda (3 |\phi|_H^2 -v^2)\Big]\Bigg\} \\
B_{H1}&=&8\pi Gr_H\left[\frac{|\Lambda|}{4\pi G} 
     -\Bigl(\frac{n^2}{r_H^2}|\phi|^2_H
     +\frac{\lambda}{2}(|\phi|^2_H-v^2)^2\Bigr)\right]
\label{b1}\\
B_{H2}&=&4\pi G \bigg[
\frac{|\Lambda|}{4\pi G} 
+\Big( \frac{n^2}{r_H^2}
-\frac{\lambda}{2}(|\phi|^2_H -v^2)^2\Big)\bigg] \nonumber\\  
&&-\frac{3|\phi|_H^2 \Big( \frac{n^2}{r_H^2}+\lambda (|\phi|_H^2-v^2)
\Big)^2 }{ 2 \Big[\frac{|\Lambda|}{4\pi G}-
(\frac{n^2}{r_H^2}|\phi|_H^2 +\frac{\lambda}{2}(|\phi|_H^2-v^2)^2) \Big] }
\label{b2}.
\end{eqnarray}
Noticing that the expansion coefficients depend only on $|\phi|_H$ 
and not on $N_H$ in Eq.~(\ref{nrh}), one may suspect a possibility that
there does not exist a smooth solution 
interpolating $|\phi|(0)=0$ and $|\phi|(\infty)=v$ 
with one horizon or two.
However, the horizon $r_H$ in the expressions of the coefficients 
is also an undetermined parameter, which is the position to which the
boundary conditions are applied.
Therefore, we can expect a difficulty to analyze the solutions 
by using the numerical technique, {\it e.g.}, the shooting method.

If $B(r)$ is a decreasing function near the origin, 
there must exist a minimum point $r_m$ of 
metric function: $B(r)\ge B_m=B(r_m)$. 
Let us do a series expansion of $B(r)$ about
the minimum
\begin{eqnarray}
|\phi|(r)&\approx& |\phi|_m +\frac{d|\phi|}{dr}\bigg|_m(r-r_m)
              +\frac{1}{2}\frac{d^2|\phi|}{dr^2}\bigg|_m(r-r_m)^2\\
N(r)&\approx& N_m 
 +8\pi G r_m \Big(\frac{d|\phi|}{dr}\bigg|_{m}\big)^2(r-r_m)
 +N_{m2} (r-r_m)^2  \\       
B(r)&\approx& B_{m} +B_{m2}(r-r_m)^2,
\end{eqnarray}
where the coefficients $\frac{d^2|\phi|}{dr^2}\Big|_m$,
$N_{m2}$,
$B_m$ and $B_{m2}$ can easily be evaluated in terms of the 
scalar amplitude $|\phi|_m$ and its derivative $\frac{d|\phi|}{dr}\Big|_m$ 
at the minimum point $r_m$:
\begin{eqnarray}
\frac{d^2|\phi|}{dr^2}\bigg|_m&=&-\bigg(8\pi Gr_{m}
\Big(\frac{d|\phi|}{dr}\bigg|_{m}\Big)^{2}+\frac{1}{r_{m}}\bigg)
\frac{d|\phi|}{dr}\bigg|_{m}\nonumber\\
&&+|\phi|_{m}\Big(\frac{d|\phi|}{dr}\bigg|_{m}\Big)^{2}
\frac{\frac{n^{2}}{r^{2}_{m}}+\lambda(|\phi|_{m}^{2}-v^{2})}{
\frac{|\Lambda|}{4\pi G}-\Big(\frac{n^{2}}{r_{m}^{2}}|\phi|^{2}_{m}
+\frac{\lambda}{2}(|\phi|^{2}_{m}-v^{2})^{2}\Big)}\\
N_{m2}&=&8\pi G \Big( \frac{d|\phi|}{dr}\bigg|_m\Big)^2\Bigg\{
-1+\frac{1}{2}\frac{d|\phi|}{dr}\bigg|_m
-8\pi G r_m^2 \Big(\frac{d|\phi|}{dr}\bigg|_m\Big)^2 \nonumber \\
&&\,\,\,+ r_m |\phi|_m \frac{d|\phi|}{dr}\bigg|_m
\frac{\frac{n^2}{r_m^2}+\lambda(|\phi|_m^2-v^2)}{
\frac{|\Lambda|}{4\pi G}-\Big(\frac{n^2}{r_m^2}|\phi|_m^2
+\frac{\lambda}{2}(|\phi|^2_m-v^2)^2} \Bigg\}\\
B_{m}&=&\frac{1}{\Big(\frac{d|\phi|}{dr}\bigg|_{m}\Big)^{2}}
\bigg[\frac{|\Lambda|}{4\pi G}-\Big(\frac{n^{2}}{r_{m}^{2}}|\phi|^{2}_{m}
+\frac{\lambda}{2}(|\phi|^{2}_{m}-v^{2})^{2}\Big)\bigg]
\label{bm1}\\
B_{m2}&=&8\pi Gr_{m}\Bigg\{-2|\phi|_{m}\frac{d|\phi|}{dr}\bigg|_{m}
\Big(\frac{n^{2}}{r_{m}^{2}}+\lambda(|\phi|^{2}_{m}-v^{2})\Big)
+\frac{n^{2}|\phi|_{m}^{2}}{r^{3}_{m}}\nonumber \\
&&\,\,\,+\bigg(8\pi Gr_{m}\Big(\frac{d|\phi|}{dr}\bigg|_{m}\Big)^{2}
+\frac{1}{r_{m}}\bigg)
\bigg[\frac{|\Lambda|}{4\pi G}-\Big(\frac{n^{2}}{r_{m}^{2}}|\phi|^{2}_{m}
+\frac{\lambda}{2}(|\phi|^{2}_{m}-v^{2})^{2}\Big)\bigg]\Bigg\}.
\end{eqnarray}

{}From now on, let us examine vortex solutions in detail for 
the cases of zero and negative cosmological constants separately.

\vskip 1em
\begin{center}{\bf A. \underline{$\Lambda=0$}}
\end{center}

First of all, under the Schwarzschild type metric in Eq.~(\ref{metric}), 
we read the case of zero cosmological 
constant. 
Consider a scalar configuration with the boundary condition $|\phi|(0)=0$,
which is consistent with the single-valuedness
of it at the origin 
under the vortex ansatz in Eq.~(\ref{ansatz}). 
Suppose that $B(r)$ is continuous and starts from $B(0)=1$.
Then, near the origin, $B(r)$ is a decreasing function as given in
Eq.~(\ref{br0}) and, since the right-hand side of Eq.~(\ref{Beq}) is negative 
as far as
$B(r)$ is positive, it is monotonically deceasing. 
{}Furthermore, since the second term of $T^{t}_{\;t}$ in
Eq.~(\ref{ttt}) is dominant for large $r$ when $|\phi|$ approaches to 
the vacuum expectation value $v$, the negativity of the coefficient of
the logarithmic term in Eq.~(\ref{binf}) tells us that $B(r)$ goes to the
negative infinity at spatial infinity if  we keep forcing the 
boundary condition $|\phi|(\infty)=v$.
It means that there should exist $r_{H}$ such that $B(r_{H})=0$. 
Obviously, the nonsingular global string solution connecting
$|\phi|(0)=0$ and $|\phi|(\infty)=v$ can not be supported 
unless the asymptotic region of
space constructed by the $U(1)$ global string is not well-behaved.
{}Furthermore, the spacetime formed by these global strings includes
a physical curvature singularity which is not removable through
coordinate transformation~\cite{Gre}.
In a viewpoint of smooth solution, there still remains
possibility of the existence of scalar
configuration which has oscillatory behavior around the vacuum expectation
value $v$ in the asymptotic region of negative $B(r)$, because the sign change 
of $B(r)$ effectively flips
up the potential term in the right-hand side of Eq.~(\ref{feq}).

\vskip 1em 
\begin{center}{\bf B. \underline{$\Lambda <0$}}
\end{center}

When the negative cosmological constant is turned on, $B(r)$ in Eq.~(\ref{br0})
near the origin increases or decreases according as its rescaled
magnitude of the cosmological constant ${|\Lambda|}/{\lambda v^{2}}$ 
whether it is larger or smaller than
the ratio of the Planck scale and the symmetry breaking scale $Gv^{2}$. Once
$B(r)$ begins with decreasing from $r=0$, sometimes it can form the horizon as
expressed in Eq.~(\ref{brh}). Here we study the vortex solutions for three
cases such that they have zero(regular vortex with asymptotic hyperbola), 
one(extremal black hole) and two(Reissner-Nordstr\"om type black hole) 
horizon(s).

\noindent B-I. \underline{Regular Neutral Vortex}:

In the previous subsection, we studied the global vortex case of zero
cosmological constant.
As expressed in Eq.~(\ref{breq}), the gravitational correction due to the
object contributes negatively to $B(r)$ and becomes dominant for large $r$ and
finally changes the sign of $B(r)$ for some $r$. 
The problem was started by
the very coordinate singularity, that is the zero point of $B(r)$, and finally
turns out to have a physical singularity.
Therefore, if the sign of $B(r)$ does not change for every $r$, there may
probably be regular global vortex solution interpolating $|\phi|(0)=0$ and
$|\phi|(\infty)=v$ smoothly.
In this viewpoint, the positive contribution of negative cosmological constant
in Eq.~(\ref{breq}) is drastic.
No matter how small the magnitude of the cosmological constant is, 
this term is dominant for
sufficiently large $r$ since it is proportional to the area of space
($\sim r^2$) for any solution with a boundary condition
${d|\phi|}/{dr}|_{r\rightarrow\infty}=0$.
In accord with another boundary condition $B(0)=1$, $B(r)$ can always
be positive for every $r$ if one adds the contribution from the first term in
Eq.~(\ref{breq}) proportional to the cosmological constant is larger than that
from the second term in Eq.~(\ref{breq}) proportional to the Newton constant.
Once $B(r)$ is regular and has only positive sign for every $r$, 
regular global vortex solution connecting  $|\phi|(0)=0$ and 
$|\phi|(\infty)=v$ smoothly, and so does $N(r)$ from Eq.~(\ref{Neq}).
The shapes of solutions obtained numerically are shown in Figure 2.
The obtained regular neutral vortex solutions of topological charge $n$ are
classified into two categories by the behavior of $B(r)$.
When $|\Lambda|>2\pi G \lambda v^4-8\pi G\phi_0^2\delta_{n1}$ in 
Eq.~(\ref{br0}), $B(r)$ is monotonically increasing as shown by the dashed
lines in Figure 2. 
When $|\Lambda|<2\pi G \lambda v^4-8\pi G\phi_0^2\delta_{n1}$  in 
Eq.~(\ref{br0}), $B(r)$ must have a positive minimum for the regular solution
as shown in the solid lines in Figure 2.

{}From Eq.~(\ref{bm1}), we can roughly 
estimate a criterion for positive $B_{m}$ or
equivalently for the regular solution:
\begin{eqnarray}
\frac{|\Lambda|}{4\pi G}
&>&\lambda\bigg[\frac{n^{2}}{(\sqrt{\lambda}vr_{m})^{2}}
\Big(\frac{|\phi|_{m}}{v}\Big)^{2}+\frac{1}{2}\bigg(\Big(\frac{|\phi|_{m}}{v}
\Big)^{2}-1\bigg)^{2}\bigg]\\
&\stackrel{\sqrt{\lambda}vr_{m}\sim 1,\;|\phi|_{m}\sim v}{\sim}
&{\cal O}(\lambda).\nonumber
\end{eqnarray}

\begin{figure}

\setlength{\unitlength}{0.1bp}
\begin{picture}(3600,2160)(0,0)
\put(3287,2060){\makebox(0,0)[l]{1.5}}
\put(3287,1423){\makebox(0,0)[l]{1}}
\put(3287,787){\makebox(0,0)[l]{0.5}}
\put(3287,150){\makebox(0,0)[l]{0}}
\put(2812,50){\makebox(0,0){6}}
\put(1962,50){\makebox(0,0){4}}
\put(1113,50){\makebox(0,0){2}}
\put(263,50){\makebox(0,0){0}}
\put(213,2060){\makebox(0,0)[r]{3}}
\put(213,1423){\makebox(0,0)[r]{2}}
\put(213,787){\makebox(0,0)[r]{1}}
\put(213,150){\makebox(0,0)[r]{0}}
\put(0,1100){\makebox(0,0)[l]{\large$B$}}
\put(1400,-100){\makebox(0,0)[l]{\large$\sqrt\lambda v r$}}
\put(3487,1100){\makebox(0,0)[l]{\large$\displaystyle\frac{|\phi|}{v}$}}
\end{picture}

\vspace{6mm}

\caption{Regular vortex configurations : $|\Lambda|/\lambda v^{2}=1.0$ and 
$8\pi Gv^{2}=1.25$ for solid lines, and $|\Lambda|/\lambda v^{2}=0.1$ and 
$8\pi Gv^{2}=1.15$ for dashed lines.}
\label{fig2}
\end{figure}
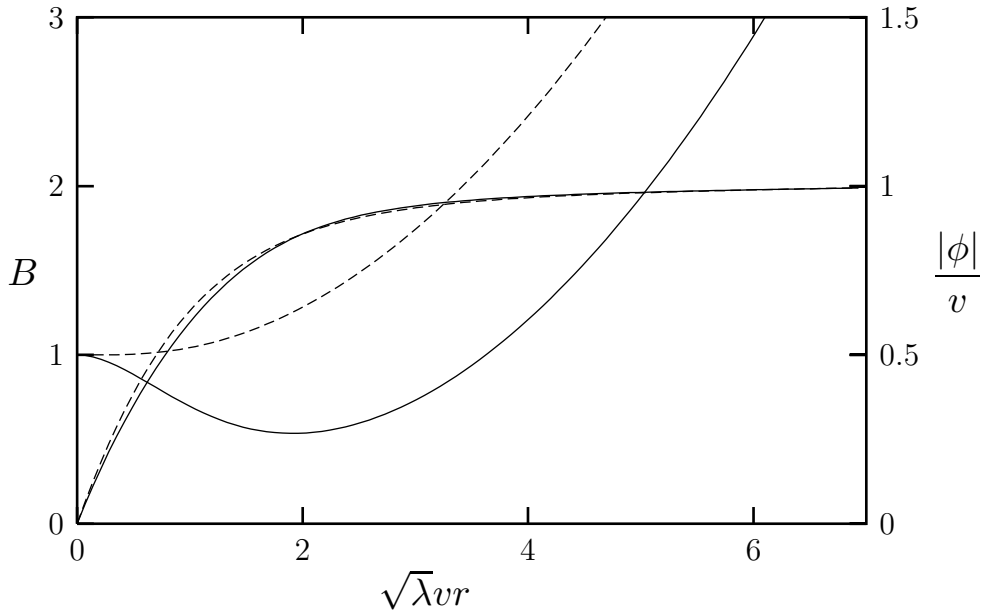

\noindent B-II. \underline{Extremal Black Hole}:

The extremal black hole has to satisfy the boundary condition 
$\frac{dB}{dr}\Big|_{r_{H}}=0$ in addition to $B(r_{H})=0$. 
Suppose that there is
an extremal black hole, the position 
of the horizon can exactly be expressed by $|\phi|_{H}$ only:
\begin{equation}
r_{H}=
\left\{\begin{array}{ll}
\sqrt{\frac{4\pi G n^2 |\phi|^2_{H}}{|\Lambda|-2\pi 
G\lambda(|\phi|^2_{H}-v^2)^2}} 
     & \mbox{from Eq.~(\ref{breq}) or}\;B_{H1}=B_{m}=0\\
\frac{n}{\sqrt{\lambda(v^2-|\phi|^2_{H}})} 
     & \mbox{from Eq.~(\ref{feq})}.
\end{array}\right.
\end{equation}
Combining these two, we obtain explicit value for $|\phi|_{H}$
\begin{eqnarray}\label{phor}
|\phi|_{H}=v\bigg(1 -\frac{|\Lambda| }{2\pi G \lambda v^4}\bigg)^{1/4}
<v,
\end{eqnarray}
and $r_H$: 
\begin{eqnarray}\label{rhor}
r_H= \frac{n}{\sqrt{\lambda}v\left(1-\sqrt{1-\frac{|\Lambda|}{2\pi
G\lambda v^4}}\right)^{1/2}}
\approx 
\left\{\begin{array}{ll}
\frac{n}{\sqrt{\lambda}v}, & \mbox{when} ~~{2\pi Gv^{2}}\approx
\frac{|\Lambda|}{\lambda v^{2}}\\
\sqrt{\frac{4\pi G\lambda v^{4}}{|\Lambda|}}\frac{n}{\sqrt{\lambda}v}, &
\mbox{when} ~~{2\pi Gv^{2}}\gg
\frac{|\Lambda|}{\lambda v^{2}}.
\end{array}\right.
\end{eqnarray}
The second line in Eq.~(\ref{rhor}) coincides with the result
in Eq.~(\ref{roughrm}) based on a rough estimation.
The complete determination of the position of the horizon $r_H$ and the
value of scalar amplitude $|\phi |_H$ is an inevitable result since the 
leading terms of Eq.~(\ref{feq}) and Eq.~(\ref{Beq}) lead to two algebraic
equations in case of the extremal black hole. Therefore, the series 
expansion of the equations (\ref{feq}), (\ref{Neq}), (\ref{Beq}) in order of
$(r-r_H)$ determines all coefficients of power series solutions $|\phi |(r)$,
$N(r)$, $B(r)$ in closed form.

Before arguing the existence of global vortex solution constituting an
extremal black hole, a comment should be placed here. 
To guarantee a solution of Eq.~(\ref{phor}) and Eq.~(\ref{rhor})
we must impose the following condition: ${|\Lambda|}/{2\pi G\lambda v^{4}}<1$.
Since this is nothing but the condition to distinguish the $n>2$ solutions
with a minimum at nonzero $r$ from those with monotonically increasing $B(r)$
in the subsection B-I. So, it is not useful except for $n=1$ solutions.
Since ${dB}/{dr}|_{H}=0$ at the horizon, $B_{H1}$ in Eq.~(\ref{brh})
and Eq.~(\ref{b1}) must vanish at the horizon $r_{H}$, and then $B_{H2}$ in
Eq.~(\ref{b2}) is automatically positive:
\begin{eqnarray}\label{b22}
B_{H2}=B_{m2}=8\pi G\lambda v^{4}\sqrt{1-\frac{|\Lambda|}{2\pi G\lambda v^{4}}}
\left(1-\sqrt{1-\frac{|\Lambda|}{2\pi G\lambda v^{4}}}\,\right) > 0 .
\end{eqnarray}
This implies convex down property of $B(r)$ at the horizon of the extremal
black hole. In addition the positivity of the slope of the scalar amplitude
gives a condition :
\begin{eqnarray}\label{cond}
\frac{|\Lambda|}{\lambda v^2} + \frac{1}{32 \pi G v^2} < \frac{1}{2},
\end{eqnarray}
which is more restrictive than the previous one from Eq.~(\ref{phor}) and
Eq.~(\ref{rhor}). The power series expansion of higher order terms will make
the condition more and more restrictive. Note that the data for our
numerical solution in the caption of Fig.~3 is consistent with 
Eq.~(\ref{cond}).

Now the remaining task is to find numerically 
a patch of solution 
connecting smoothly the boundary conditions $(|\phi|(0)=0, B(0)=1)$
and $(|\phi|_{H}=(\ref{phor}), B_{H}={dB}/{dr}|_{H}=0)$
by use of the shooting method for various $(\phi_{0}, N_{0})$'s in 
Eq.~(\ref{phi00}) and Eq.~(\ref{br0}), 
and the other patch from the horizon to spatial infinity 
$(|\phi|(\infty)=v, N(\infty)=1)$. 
We obtain one for a model with 
$8\pi Gv^2=1.338$ and ${|\Lambda |}/{\lambda v^2}=0.1$ (See Figure 3).

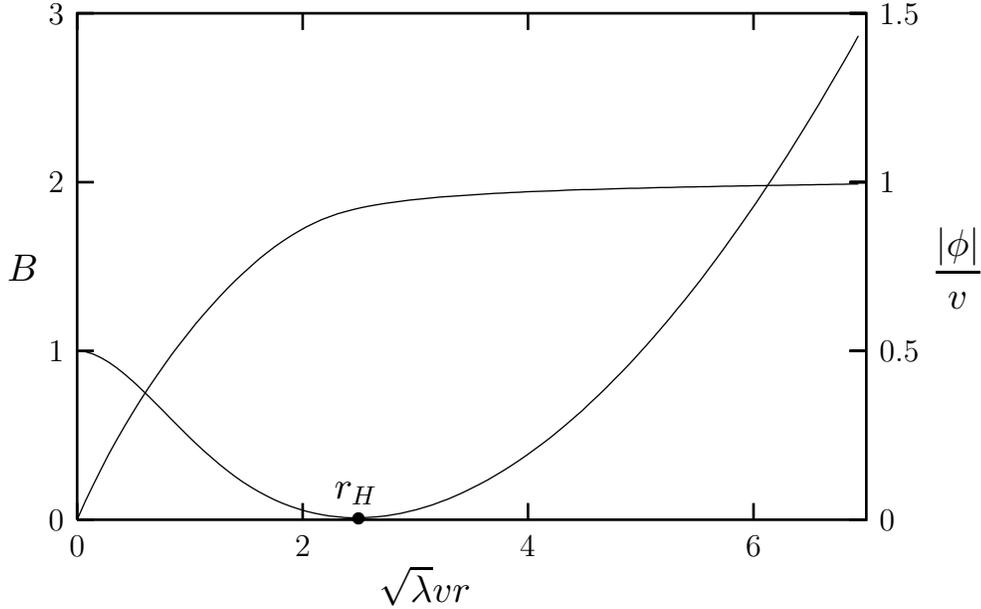
\begin{figure}

\setlength{\unitlength}{0.1bp}
\begin{picture}(3600,2160)(0,0)
\put(1393,250){\makebox(0,0)[r]{\large$r_H$}}
\put(1353,150){\makebox(0,0)[r]{$\bullet$}}
\put(3287,2060){\makebox(0,0)[l]{1.5}}
\put(3287,1423){\makebox(0,0)[l]{1}}
\put(3287,787){\makebox(0,0)[l]{0.5}}
\put(3287,150){\makebox(0,0)[l]{0}}
\put(2812,50){\makebox(0,0){6}}
\put(1962,50){\makebox(0,0){4}}
\put(1113,50){\makebox(0,0){2}}
\put(263,50){\makebox(0,0){0}}
\put(213,2060){\makebox(0,0)[r]{3}}
\put(213,1423){\makebox(0,0)[r]{2}}
\put(213,787){\makebox(0,0)[r]{1}}
\put(213,150){\makebox(0,0)[r]{0}}
\put(0,1100){\makebox(0,0)[l]{\large$B$}}
\put(1400,-100){\makebox(0,0)[l]{\large$\sqrt\lambda v r$}}
\put(3487,1100){\makebox(0,0)[l]{\large$\displaystyle\frac{|\phi|}{v}$}}
\end{picture}

\vskip 6mm 
\caption{Extremal black hole solution. 
$|\Lambda|/\lambda v^2=0.1$ and $8\pi Gv^2=1.338$.}
\label{fig3}
\end{figure}

\noindent B-III. \underline{Charged Black Hole with Two Horizons}:

As the scale of cosmological constant $|\Lambda|/\lambda v^{2}$ becomes
smaller than the critical value to support the extremal black hole, 
we expect to witness a charged black hole with two horizons.
One can easily read this phenomenon by examining 
the integral equation of $B(r)$. 
The right-hand side of Eq.~(\ref{breq}) involves 
two contributions; positive term is proportional to $|\Lambda|$ 
while negative one is proportional to $G$. 
Obviously two terms are zero at the origin, so $B(0)$ is equal to one
(positive). If the second term dominates due to 
${|\Lambda|}/{\lambda v^{2}}\ll 8\pi Gv^{2}$, 
then $B(r)$ becomes negative at some intermediate
region. However, for sufficiently large $r$, the first term proportional to
the square of the radius $r$ is much larger than the second term of which the
leading contribution is logarithmic, and finally $B(r)$ becomes positive
again. 

{}From now on, let us discuss details about the existence of two horizons in
several steps. The first step is to show that there exists the 
inner horizon
$r_{in}$ where $|\phi|(r_{in})<v$ and ${d|\phi|}/{dr}|_{r_{H}}>0$. 
Let us assume a situation that 
$2\pi G(\lambda v^{4}+4\phi^{2}_{0}\delta_{1n})$ is
much larger than $|\Lambda|$ in Eq.~(\ref{br0}) and
$|\phi|(r)\approx\phi_{0}r^{n}$. Then there exists an appropriate $\phi_{0}$
for a sufficiently small $r$ such that $|\phi|(r)<v$ and $B(r)$ hits the zero
point approximately at 
$r_{in}\approx 1/\sqrt{2\pi G(\lambda
v^{4}+4\phi^{2}_{0}\delta_{1n})-|\Lambda|}$. 
The second step deals the outer
horizon. Again let us assume the scaled cosmological constant
$|\Lambda|/\lambda v^{2}$ is much smaller than the ratio of the square of
symmetry breaking scale and that of the Planck scale $2\pi Gv^{2}$. Then there
exists obviously a position $r$ such that Eq.~(\ref{binf}) without the term of
order $1/r^{2}$ is valid $(r\gg r_{c})$ and, simultaneously,
$|\Lambda|r^{2}$ term can also be neglected:
\begin{eqnarray}
(\ref{binf})&\sim&-8\pi Gv^{2}n^{2}\ln r/r_{c}-8G{\cal M}+1\nonumber\\
&\sim&4\pi G v^2 n^2\Big(1-2\ln r/r_{c}\Big)+1.\label{besti}
\end{eqnarray} 
If the second logarithmic term in the right-hand side of
Eq.~(\ref{besti}) is larger than order one for some $r$, the value of $B(r)$
is negative at this range of $r$. For an arbitrarily small positive
$\phi_{\infty}$, $|\phi|(r)<v$. In addition, $\frac{d|\phi|}{dr}\Big|_{H}$ is
positive because of the negativity of both numerator and denominator in
Eq.~(\ref{dfdr}). 
The third step begins with recalling  $r_m$ obtained in Eq.~(\ref{roughrm})  
by minimizing $B(r)$.
{}For the condition $2\pi Gv^2 > |\Lambda|/\lambda v^2$, 
$r_m>r_H^{in}$. 
Since Eq.~(\ref{binf}) tells $B(r_m)<0$ when $8\pi Gv^2 n^2 >1$,
$r_{out}$ should exist and be larger than $r_m$. 
Therefore, the remaining step is to find a smooth scalar field
$|\phi|(r)$ to have two horizons $r_H^{in}$ and $r_H^{out}$ through
numerical analysis.
The above range does not forbid the convexity of the metric function $B(r)$
both for the shooting from the outside and for the shooting from the inside, 
it implies the possibility of the existence of a smooth configuration to
connect $|\phi|(0)=0$ and $|\phi|(\infty)=v$. The approximate
value of the inner horizon is $\sqrt{\lambda}vr_{H}^{in}\approx 2.03 $ 
and that of the outer
horizon is $\sqrt{\lambda} vr_{H}^{out} \approx 3.05 $ from Fig.~4.

\begin{figure}

\setlength{\unitlength}{0.1bp}
\begin{picture}(3600,2160)(0,0)
\put(1143,423){\makebox(0,0){$\bullet$}}
\put(1143,573){\makebox(0,0){\large$r_H^{in}$}}
\put(1553,423){\makebox(0,0){$\bullet$}}
\put(1553,573){\makebox(0,0){\large$r_H^{out}$}}
\put(3287,2060){\makebox(0,0)[l]{1.5}}
\put(3287,1514){\makebox(0,0)[l]{1}}
\put(3287,969){\makebox(0,0)[l]{0.5}}
\put(3287,423){\makebox(0,0)[l]{0}}
\put(2812,50){\makebox(0,0){6}}
\put(1962,50){\makebox(0,0){4}}
\put(1113,50){\makebox(0,0){2}}
\put(263,50){\makebox(0,0){0}}
\put(213,2060){\makebox(0,0)[r]{3}}
\put(213,1514){\makebox(0,0)[r]{2}}
\put(213,969){\makebox(0,0)[r]{1}}
\put(213,423){\makebox(0,0)[r]{0}}
\put(0,1200){\makebox(0,0)[l]{\large$B$}}
\put(1400,-100){\makebox(0,0)[l]{\large$\sqrt\lambda v r$}}
\put(3487,1200){\makebox(0,0)[l]{\large$\displaystyle\frac{|\phi|}{v}$}}
\end{picture}

\vskip 6mm
\caption{A charged black hole solutions with two horizons
$ r_H^{in}$ and $r_H^{out}$. $|\Lambda|/\lambda v^2=0.1$ and $8\pi G
v^2=1.4$.}
\label{fig4}
\end{figure}
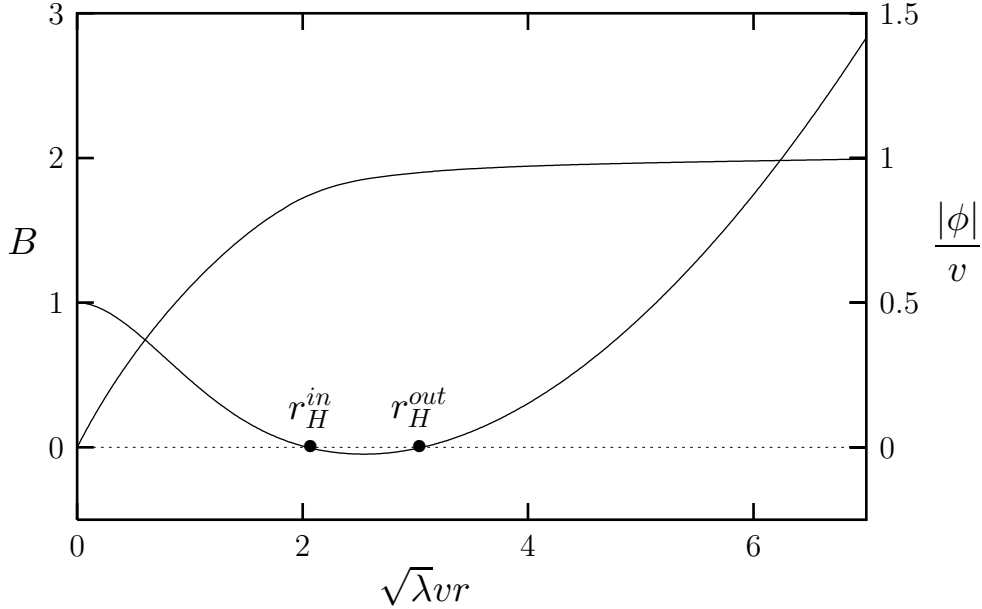

Now we have three types of global $U(1)$ vortices in anti-de Sitter 
spacetime. 
Smooth as the configurations of scalar field $|\phi|(r)$ are,
the forms of metric function $B(r)$ contain horizons.
At this stage, we should make it clear whether the singularity at each
horizon is a coordinate artifact or a physical singularity. The answer is given
by examining the square of the the curvature: Specifically, the Kretschmann
scalar is expressed by the Einstein tensors in (2+1) dimensions 
\begin{eqnarray}
\lefteqn{R_{\mu\nu\rho\sigma}R^{\mu\nu\rho\sigma}}&&\nonumber\\
&=& 4G_{\mu\nu}G^{\mu\nu}\label{curv}\\
&=& 4\mbox{Tr}
\bigg[\mbox{diag}\Big(-\frac{1}{2r}\frac{dB}{dr},-\frac{1}{2r}\frac{dB}{dr}
-\frac{B}{r}\frac{dN}{dr}, -\frac{1}{2}\frac{d^{2}B}{dr^{2}}-\frac{3}{2}
\frac{dB}{dr}\frac{dN}{dr}-B\frac{d^{2}N}{dr^{2}}-B\Big(\frac{dN}{dr}\Big)^{2}
\Big)\bigg].\nonumber 
\end{eqnarray}
At first glance one can read no curvature singularity at the horizons
from the expression Eq.~(\ref{curv}). The forms of Eq.~(\ref{Neq}) and
Eq.~(\ref{Beq}) tell us that there is no divergent curvature at the origin
as far as the scalar field $|\phi|(r)$ behaves regularly, i.e., $|\phi|(r)
\sim r^{n}$.  

\vspace{5mm}
 
As mentioned previously, a 
characteristic of BTZ black holes is that
they need not contain the divergent
curvature at real $r$. We have understood that the
Schwarzschild type solution 
in Eq.~(\ref{Schw}) is regular everywhere and the charged
black hole formed by the natural vortex discussed 
in the subsection B-III has  also a
divergent curvature counterpart of 
charged BTZ black hole formed by the electric point source
in Eq.~(\ref{RNbh}). 
Here let us clarify the structure of manifolds and the nature of forces due to
the global vortices by studying the geodesics of test particles~\cite{CMP}.

The geometry depicted by the metric in Eq.~(\ref{demet})
admits two Killing vectors, $\partial/\partial t $ and
$\partial/\partial\theta$, and then the 
constants of motion along the geodesic are
\begin{eqnarray}
\gamma= B e^{2N} \frac{dt}{ds}~~{\rm and}~~
L= r^2\frac{d\theta}{ds},
\end{eqnarray}
where $s$ is an affine parameter along the geodesic.
Note that one can not interpret
$\gamma$ as the local energy of a particle at the spatial infinity,
since the spacetime is not asymptotically flat.
Using these constants of motion, 
we obtain the following geodesic equation for the radial motions:
\begin{eqnarray}
\label{radial}
\frac{1}{2}\Big(\frac{dr}{ds}\Big)^2&=& -\frac{1}{2}\bigg(
B(r)\Big(m^2 +\frac{L^2}{r^2}\Big)
-\frac{\gamma^2}{e^{2N(r)}}\bigg)=-V(r).
\end{eqnarray}
where $m$ can be rescaled to be one  for any time-like geodesic 
and zero for a null-like geodesic.
The radial equation in Eq.~(\ref{radial}) is an analogue of Newton's
equation for $r\ge0$ with conservative effective potential
$V(r)$
in which the hypothetical particle has unit mass and zero total
mechanical energy.
In order to identify the existence of the black hole, 
a meaningful quantity is 
the elapsed coordinate time $t$ of the static observer at $r_{0}$ for the 
motion of a test particle from $r_{0}$ to $r$:
\begin{eqnarray}\label{statict}
t=\int^{r_0}_{r}\frac{dr}{B(r)e^{N(r)}
\sqrt{1-\frac{1}{\gamma^2}(m^{2}+\frac{L^2}{r^{2}})
B(r)e^{2N(r)}}}.
\end{eqnarray}
 
When we analyze the motions of test particles, they are divided into four
categories that whether they have mass $(m=1)$ or not $(m=0)$, or whether
their motions are purely radial $(L=0)$ or rotating $(L\ne 0)$.
{} For simplicity, we use the numbering of subsections in accord with the
previous ones for vortex solutions;
B-I for regular solution, B-II for extremal black hole and B-III
for charged black hole.

B-I-(a) \underline{$(m=0,L=0)$}: For the radial motion of a massless test
particle, the effective potential has no explicit dependence on $B(r)$ such
that
\begin{eqnarray}\label{pot00}
V(r)=-\frac{\gamma^{2}}{2}e^{-2N(r)}\leq 0.
\end{eqnarray}
The allowed motions are (i) stopped particle motion for $\gamma=0$,
 which is unstable,
and (ii) an unbounded motion for $\gamma\neq 0$  
since the potential is negative everywhere even at spatial infinity. 
Moreover, $N(r)$ is monotonic increasing, 
so the radial force is always attractive. Then
the test particle starts with initial speed greater than 
${dr}/{d\tau}={\gamma}/{\sqrt{2}}$ 
at a point $r_{0}$ and approaches to the center of 
the vortex. Obviously there is no horizon $(B(r)>0$ for all $r$), and the test
particle started from a finite initial position $r_{0}$ arrives at the origin
in  a finite time measured by the clock of the static observer, i.e.,
$t=\int^{r_0}_{0}{dr}/{B(r)e^{N(r)}}$ is finite. 

B-I-(b) \underline{$(m=0,L\ne 0)$}: For a rotational motion, the centrifugal
force term is introduced in the potential $V(r)$:
\begin{eqnarray}\label{pot01}
V(r)=\frac{1}{2}\bigg(\frac{L^{2}B(r)}{r^{2}}-\frac{\gamma^{2}}{e^{2N(r)}}
\bigg) 
\end{eqnarray}
which forbids the motion near the core of the vortex. 
{}From Figure 5,
we read the minimum value of $\gamma$, $\gamma_{cr}$, which means no motion is
allowed for $\gamma$ smaller than $\gamma_{cr}$ (See the dotted line in Fig.~5)
and the circular orbit is at the radius $r_{cr}$ for $\gamma=\gamma_{cr}$
(See the solid line in Fig.~5). When $\gamma>\gamma_{cr}$, the motions are
classified into two categories according to the values of $\gamma/L$: 
(i) When $\gamma/L\geq\sqrt{|\Lambda|}$, the motion is unbounded in which the
speed of the test particle at spatial infinity is 
${dr}/{d\tau}|_{r=\infty}=\sqrt{L^{2}|\Lambda|-\gamma^{2}}$ (See dashed 
lines (ii) and (iii) in Fig.~5).
(ii) When $\gamma_{cr}/L<\gamma/L<\sqrt{|\Lambda|}$, it is a bounded orbit with
perihelion $r_{min}$ and aphelion $r_{max}$ (See a dashed line (i) in Fig.~5).
For the bounded motion, the apsidal distance $r_{max}-r_{min}$ is in order
$\sqrt\lambda v$.

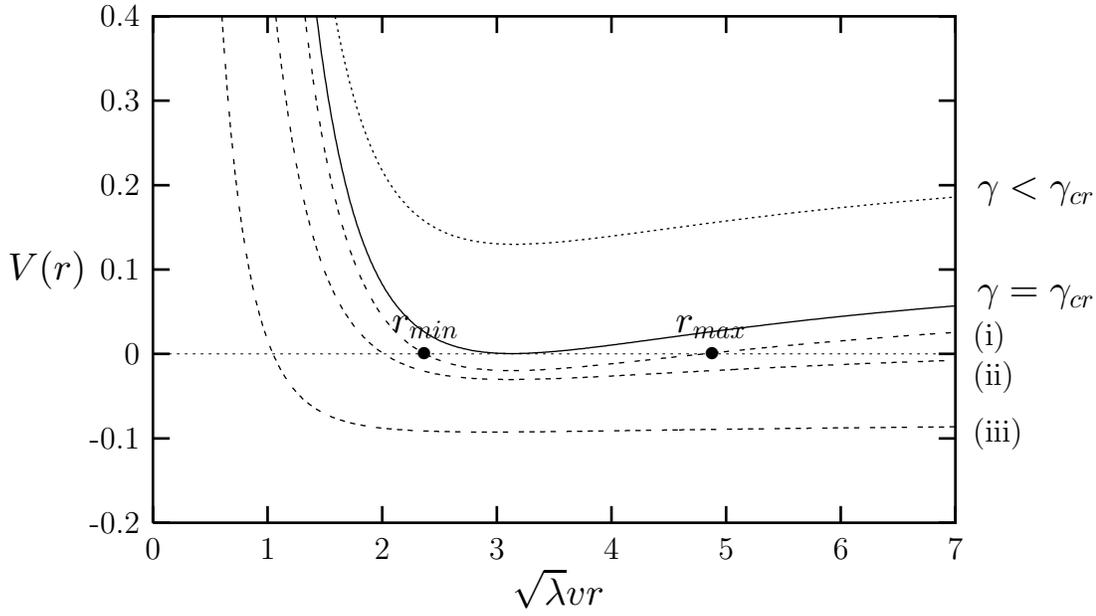
\begin{figure}

\setlength{\unitlength}{0.1bp}
\begin{picture}(3600,2160)(0,0)
\put(1941,-100){\makebox(0,0){\large $\sqrt\lambda v r $}}
\put(163,1105){\makebox(0,0)[r]{\large $V(r)$}}
\put(3963,1400){\makebox(0,0)[r]{\large$\gamma < \gamma_{cr}$}}
\put(3963,1000){\makebox(0,0)[r]{\large$\gamma = \gamma_{cr}$}}
\put(3503,850){\makebox(0,0)[l]{(i)}}
\put(3503,700){\makebox(0,0)[l]{(ii)}}
\put(3503,486){\makebox(0,0)[l]{(iii)}}
\put(1465,787){\makebox(0,0)[r]{$\bullet$}}
\put(1565,887){\makebox(0,0)[r]{\large$r_{min}$}}
\put(2550,787){\makebox(0,0)[r]{$\bullet$}}
\put(2650,887){\makebox(0,0)[r]{\large$r_{max}$}}
\put(3437,50){\makebox(0,0){7}}
\put(3005,50){\makebox(0,0){6}}
\put(2573,50){\makebox(0,0){5}}
\put(2141,50){\makebox(0,0){4}}
\put(1709,50){\makebox(0,0){3}}
\put(1277,50){\makebox(0,0){2}}
\put(845,50){\makebox(0,0){1}}
\put(413,50){\makebox(0,0){0}}
\put(363,2060){\makebox(0,0)[r]{0.4}}
\put(363,1742){\makebox(0,0)[r]{0.3}}
\put(363,1423){\makebox(0,0)[r]{0.2}}
\put(363,1105){\makebox(0,0)[r]{0.1}}
\put(363,787){\makebox(0,0)[r]{0}}
\put(363,468){\makebox(0,0)[r]{-0.1}}
\put(363,150){\makebox(0,0)[r]{-0.2}}
\end{picture}

\vskip 2em 
\caption{Schematic shapes of the effective potential $V(r)$ for the
rotational motions of the massless test particle when $L=1$.
Here $B(r)$ is always positive.}
\label{fig5}
\end{figure}

B-I-(c) \underline{$(m=1,L=0)$}: For the radial motion of a massive test
particle, the potential of it is
\begin{eqnarray}\label{pot10}
V(r)=\frac{1}{2}\bigg(B(r)-\frac{\gamma^{2}}{e^{2N(r)}}\bigg).
\end{eqnarray}
Since $B(r)\sim|\Lambda|r^{2}$ for sufficiently large $r$, all possible
motions are bounded, i.e., there exists $r_{max}$ such as $r\leq r_{max}$
for any position $r$ of the test particle.
When $B(r)$ is monotonic increasing, $V(r)$ is also monotonic increasing and
thereby the force is attractive everywhere. It is physically natural since
sufficiently-large negative vacuum energy can pervade all space despite the
repulsive force at the core of the vortex. Therefore, the motions are allowed
only when $\gamma\geq e^{N(0)}$: When $\gamma=e^{N(0)}$, the allowed motion is
stopped one at the origin, and, when $\gamma>e^{N(0)}$, the particle can move
inside $r_{max}$ and the stopped test particle starts to go inward.
When $B(r)$ has positive minimum at a nonzero $r$, the schematic shapes of the
effective potential $V(r)$ are given in Fig.~6, 
and the possible radial motions are
classified in Table. 1 since it seems that $\gamma_{stop}<e^{N(0)}$ in our
solution.

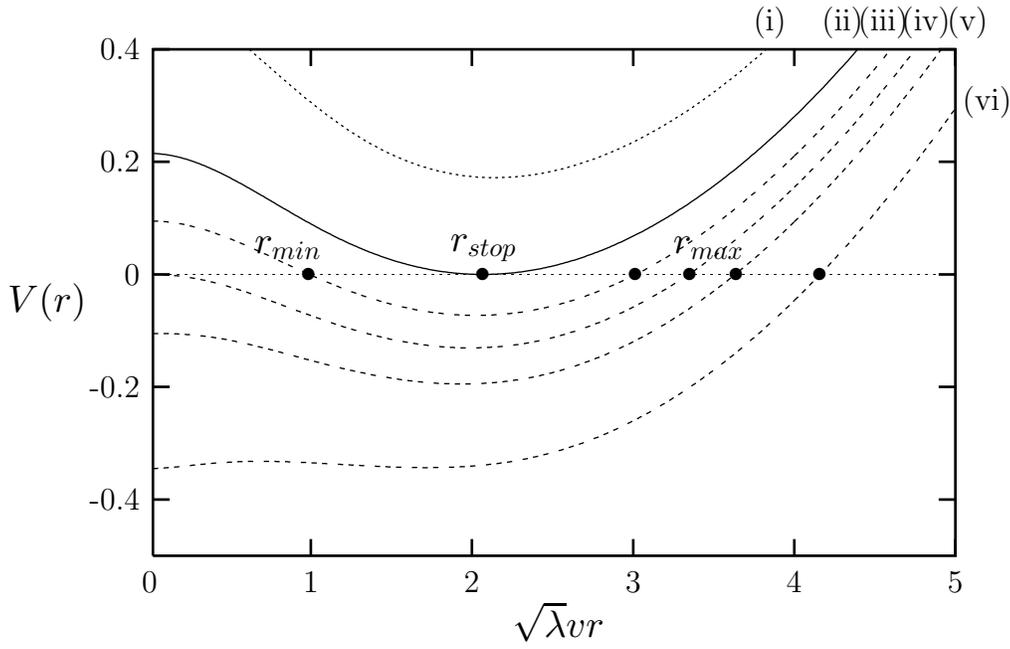
\begin{figure}

\setlength{\unitlength}{0.1bp}
\begin{picture}(3600,2160)(0,0)
\put(1941,-100){\makebox(0,0){\large $\sqrt\lambda v r $}}
\put(163,1105){\makebox(0,0)[r]{\large $V(r)$}}
\put(2800,2160){\makebox(0,0)[r]{(i)}}
\put(3090,2160){\makebox(0,0)[r]{(ii)}}
\put(3260,2160){\makebox(0,0)[r]{(iii)}}
\put(3420,2160){\makebox(0,0)[r]{(iv)}}
\put(3560,2160){\makebox(0,0)[r]{(v)}}
\put(3650,1860){\makebox(0,0)[r]{(vi)}}
\put(1685,1211){\makebox(0,0)[r]{$\bullet$}}
\put(1785,1311){\makebox(0,0)[r]{\large$r_{stop}$}}
\put(1028,1211){\makebox(0,0)[r]{$\bullet$}}
\put(1048,1311){\makebox(0,0)[r]{\large$r_{min}$}}
\put(2260,1211){\makebox(0,0)[r]{$\bullet$}}
\put(2465,1211){\makebox(0,0)[r]{$\bullet$}}
\put(2640,1211){\makebox(0,0)[r]{$\bullet$}}
\put(2640,1311){\makebox(0,0)[r]{\large$r_{max}$}}
\put(2955,1211){\makebox(0,0)[r]{$\bullet$}}
\put(3437,50){\makebox(0,0){5}}
\put(2830,50){\makebox(0,0){4}}
\put(2223,50){\makebox(0,0){3}}
\put(1615,50){\makebox(0,0){2}}
\put(1008,50){\makebox(0,0){1}}
\put(400,50){\makebox(0,0){0}}
\put(363,2060){\makebox(0,0)[r]{0.4}}
\put(363,1636){\makebox(0,0)[r]{0.2}}
\put(363,1211){\makebox(0,0)[r]{0}}
\put(363,787){\makebox(0,0)[r]{-0.2}}
\put(363,362){\makebox(0,0)[r]{-0.4}}
\end{picture}

\vskip 2em 
\caption{Schematic shapes of the effective potential $V(r)$ for the
radial motions of the massive particle. Here $B(r)$ has its positive minimum 
at a positive $r$.}
\label{fig6}
\end{figure}

\begin{center}{
\begin{tabular}{|c|c|c|} \hline
Fig.~6 & $\gamma$ & orbit,$\;$force\\ \hline
(i) & $0\leq\gamma<\gamma_{stop}$ & no orbit \\ \hline
(ii) & $\gamma=\gamma_{stop}$ & stopped motion at $r_{stop}$ \\ \hline
(iii) & $\gamma_{stop}<\gamma<e^{N(0)}$ & oscillation between $r_{min}$ and
$r_{max}$ \\ \hline
(iv) & $\gamma=e^{N(0)}$ & $r_{min}=0$ \\ \hline
(v) & $e^{N(0)}<\gamma<\gamma_{cr}$ & $r\leq r_{max}$, repulsive near the core
\\ \hline
(vi) & $\gamma\geq\gamma_{cr}$ & $r\leq r_{max}$, attractive everywhere
\\ \hline
\end{tabular}
}\end{center}

\vspace{5mm}

\noindent Table 1. The radial motions of a test particle of mass $m=1$ for
various $\gamma$'s, when $|\Lambda|/\lambda v^{2}=$1.0, $Gv^{2}=$1.0  and 
$L$ is rescaled to one.

\vspace{5mm}

B-I-(d) \underline{$(m=1,L\neq 0)$}: For the rotational motions of a massive
test particle, the effective potential takes general form
\begin{eqnarray}\label{pot11}
V(r)=\frac{1}{2}\bigg[B(r)\Big(1+\frac{L^{2}}{r^{2}}\Big)
-\frac{\gamma^{2}}{e^{2N(r)}}\bigg].
\end{eqnarray}
$V(r)\sim\frac{1}{2}|\Lambda|r^{2}$ for large $r$, and
$V(r)\sim{L^{2}}/2{r^{2}}$ for small $r$. 
Therefore, there exists a critical
value of $\gamma$, $\gamma_{circ}$, for positive $B(r)$ that there is no 
orbit for $\gamma$ smaller than this critical value $\gamma_{circ}$.
The allowed motions are (i) the circular orbit at $r_{circ}$ when $\gamma=
\gamma_{circ}$, and (ii) the bounded orbit between perihelion $r_{min}$ and
aphelion $r_{max}$ when $\gamma$ is larger than $\gamma_{circ}$ (See Fig.~7). 
Similar to the previous bounded orbit motions, the range of allowed region
is roughly estimated as a few $1/\sqrt\lambda v$.

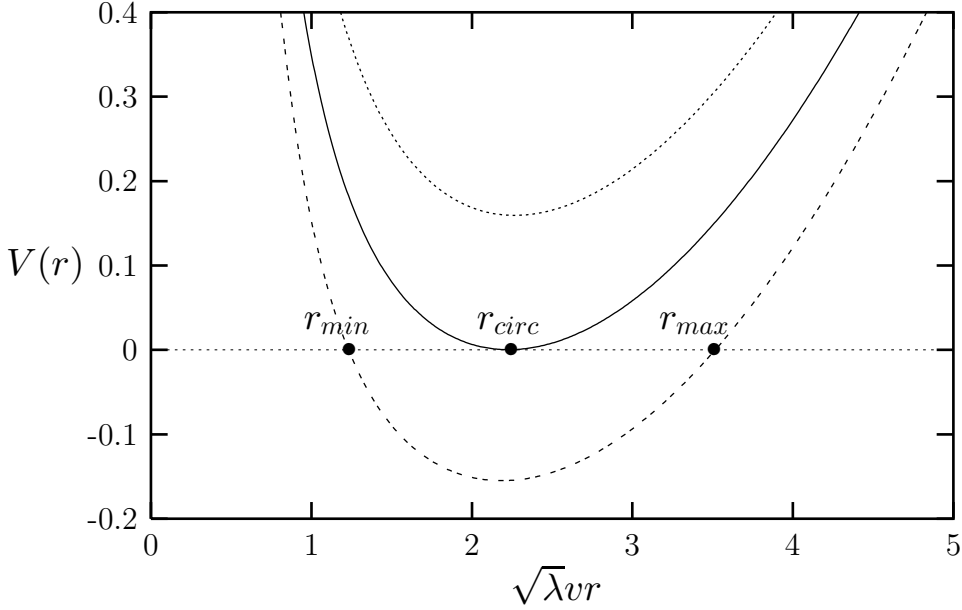
\begin{figure}

\setlength{\unitlength}{0.1bp}
\begin{picture}(3600,2160)(0,0)
\put(1941,-100){\makebox(0,0){\large $\sqrt\lambda v r $}}
\put(163,1105){\makebox(0,0)[r]{\large $V(r)$}}
\put(1800,787){\makebox(0,0)[r]{$\bullet$}}
\put(1880,887){\makebox(0,0)[r]{\large$r_{circ}$}}
\put(1190,787){\makebox(0,0)[r]{$\bullet$}}
\put(1240,887){\makebox(0,0)[r]{\large$r_{min}$}}
\put(2565,787){\makebox(0,0)[r]{$\bullet$}}
\put(2595,887){\makebox(0,0)[r]{\large$r_{max}$}}
\put(3437,50){\makebox(0,0){5}}
\put(2832,50){\makebox(0,0){4}}
\put(2227,50){\makebox(0,0){3}}
\put(1623,50){\makebox(0,0){2}}
\put(1018,50){\makebox(0,0){1}}
\put(413,50){\makebox(0,0){0}}
\put(363,2060){\makebox(0,0)[r]{0.4}}
\put(363,1742){\makebox(0,0)[r]{0.3}}
\put(363,1423){\makebox(0,0)[r]{0.2}}
\put(363,1105){\makebox(0,0)[r]{0.1}}
\put(363,787){\makebox(0,0)[r]{0}}
\put(363,468){\makebox(0,0)[r]{-0.1}}
\put(363,150){\makebox(0,0)[r]{-0.2}}
\end{picture}

\vskip 2em
\caption{Schematic shapes of the effective potential $V(r)$ for
the rotational motions of the massive particle. $\gamma_{circ}$ is 0.832
and $L=1$.}
\label{fig7}
\end{figure}

Now we have the global vortex configurations with horizons, i.e., the points 
of vanishing $B(r)$. We examine the possible motions of massless and massive
test particles under the influence of this geometry and identify these
manifolds as those of extremal and Reissner-Nordstr\"om type black holes.
Similar to the case of regular solutions, we analyze the orbits for four
categories. For the extremal case, there is no distinction from
Reissner-Nordstr\"om case when we set $r_{H}=r^{in}_{H}=r^{out}_{H}$.

B-II(III)-(a) \underline{$(m=0,L=0)$}: For the radial motions of a massless
test particle, the effective potential $V(r)$ does not depend on $B(r)$ as in
Eq.~(\ref{pot00}), so the
analysis in B-I-(a) is the same as that for this case. However, since $B(r)$
includes negative region between two horizons 
$r^{in}_{H}$ and $r^{out}_{H}$, the radial
motions are divided into two; one around the vortex core inside the inner
horizon $(r<r^{in}_{H})$ and the other outside the outer horizon
$(r>r^{out}_{H})$. Though the motions at the particle's coordinates resemble
those of regular $B(r)$, they are observed with drastic difference to the 
static observer. Since $B(r)$ vanishes both at inner and outer horizons, the
elapsed time $t$ to reach a horizon is logarithmically divergent in terms of
coordinate time for the static observer:
\begin{eqnarray}
t&\sim&\lim_{\varepsilon\rightarrow 0^{+}}\int^{r_{H}^{out}+\varepsilon}_{r_{0}}
\frac{dr}{B(r)}\\
&\sim&\bigg(\lim_{\varepsilon\rightarrow
0^{+}}\int^{r_{H}^{out}+\varepsilon}_{r_{0}}\frac{dr}{r-r^{out}_{H}}\bigg)
\times (\mbox{finite part}),
\label{cotime}
\end{eqnarray}
where $r_{0}>r_{H}^{out}+\varepsilon$. Since the potential is attractive
outside the outer horizon, the ingoing particle takes infinite time to reach
the outer horizon for the static observer. If we replace
$r_{H}^{out}+\varepsilon$ to $r_{H}^{in}-\varepsilon$ and
$r_{0}<r_{H}^{in}-\varepsilon$, then one can easily notice that the situation
is the same for the case inside the inner horizon. 
However, one must remember the attractive nature
of the force inside the inner horizon,
which causes the test particle 
to move the center of the vortex.

B-II(III)-(b) \underline{$(m=0,L\ne 0)$}: The rotational motion of a
massless test particle is described by the effective 
potential in Eq.~(\ref{pot01}).
Since $B(r)$ is negative between the horizons $(r^{in}_{H}<r<r^{out}_{H})$ and
$N(r)$ term in Eq.~(\ref{pot01}) is always negative, the shapes of the
potential for this case correspond to the dashed lines ((i), (ii), (iii)) in 
Figure 5. Then the allowed regions are as follows: When $\gamma/L<
\sqrt{|\Lambda|}$, $r_{min}\leq r<r^{in}_{H}$ and $r^{out}_{H}<r\leq r_{max}$
for two bounded motions. When $\gamma/L\geq \sqrt{|\Lambda|}$, $r_{min}\leq
r<r^{in}_{H}$ for a 
bounded motion inside the black hole and $r^{out}_{H}<r$ for
an unbounded motion outside the outer horizon. As noted 
in Eq.~(\ref{cotime}), 
the time elapsed to reach a horizon for a static observer 
is infinite, which means no orbital motion. 

B-II(III)-(c) \underline{$(m=1,L=0)$}: As discussed in B-II(III)-(b), $V(r)$
includes negative region between $r_{min}\;(r_{min}<r^{in}_{H})$ and
$r_{max}\;(r_{max}>r^{out}_{H})$. Therefore, for various $\gamma$ values, one
can expect two patterns; one is given in the lines (iii)$\sim$(vi) in Fig.~6
and the other is summarized  below (See Figure 8). 
The corresponding solutions are
provided in Table 1 for the former and in Table 2 for the latter.
For the BTZ black hole  solutions we obtained, the extremal solution follows the
Figure 6 ((iii)$\sim$(vi) in Table 1) and the BTZ solution with two horizons 
follows Figure 8 (Table 2).

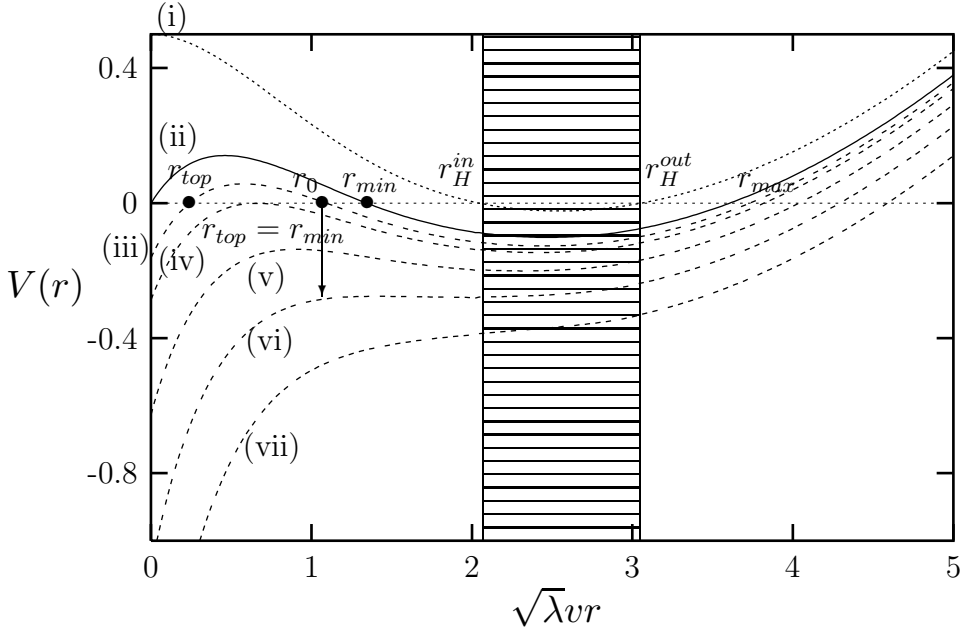
\begin{figure}

\setlength{\unitlength}{0.1bp}
\begin{picture}(3600,2160)(0,0)
\put(1941,-100){\makebox(0,0){\large $\sqrt\lambda v r $}}
\put(163,1105){\makebox(0,0)[r]{\large $V(r)$}}
\put(1665,150){\line(0,1){1910}}
\put(2257,150){\line(0,1){1910}}
\put(1665,200){\line(1,0){592}}
\put(1665,250){\line(1,0){592}}
\put(1665,300){\line(1,0){592}}
\put(1665,350){\line(1,0){592}}
\put(1665,400){\line(1,0){592}}
\put(1665,450){\line(1,0){592}}
\put(1665,500){\line(1,0){592}}
\put(1665,550){\line(1,0){592}}
\put(1665,600){\line(1,0){592}}
\put(1665,650){\line(1,0){592}}
\put(1665,700){\line(1,0){592}}
\put(1665,750){\line(1,0){592}}
\put(1665,800){\line(1,0){592}}
\put(1665,850){\line(1,0){592}}
\put(1665,900){\line(1,0){592}}
\put(1665,950){\line(1,0){592}}
\put(1665,1000){\line(1,0){592}}
\put(1665,1050){\line(1,0){592}}
\put(1665,1100){\line(1,0){592}}
\put(1665,1150){\line(1,0){592}}
\put(1665,1200){\line(1,0){592}}
\put(1665,1250){\line(1,0){592}}
\put(1665,1300){\line(1,0){592}}
\put(1665,1350){\line(1,0){592}}
\put(1665,1400){\line(1,0){592}}
\put(1665,1450){\line(1,0){592}}
\put(1665,1500){\line(1,0){592}}
\put(1665,1550){\line(1,0){592}}
\put(1665,1600){\line(1,0){592}}
\put(1665,1650){\line(1,0){592}}
\put(1665,1700){\line(1,0){592}}
\put(1665,1750){\line(1,0){592}}
\put(1665,1800){\line(1,0){592}}
\put(1665,1850){\line(1,0){592}}
\put(1665,1900){\line(1,0){592}}
\put(1665,1950){\line(1,0){592}}
\put(1665,2000){\line(1,0){592}}
\put(1665,2050){\line(1,0){592}}
\put(3437,50){\makebox(0,0){5}}
\put(2832,50){\makebox(0,0){4}}
\put(2227,50){\makebox(0,0){3}}
\put(1623,50){\makebox(0,0){2}}
\put(1018,50){\makebox(0,0){1}}
\put(413,50){\makebox(0,0){0}}
\put(363,1933){\makebox(0,0)[r]{0.4}}
\put(363,1423){\makebox(0,0)[r]{0}}
\put(363,914){\makebox(0,0)[r]{-0.4}}
\put(363,405){\makebox(0,0)[r]{-0.8}}
\put(553,2120){\makebox(0,0)[r]{(i)}}
\put(593,1670){\makebox(0,0)[r]{(ii)}}
\put(413,1250){\makebox(0,0)[r]{(iii)}}
\put(623,1200){\makebox(0,0)[r]{(iv)}}
\put(923,1140){\makebox(0,0)[r]{(v)}}
\put(953,900){\makebox(0,0)[r]{(vi)}}
\put(977,500){\makebox(0,0)[r]{(vii)}}
\put(1637,1550){\makebox(0,0)[r]{$r_H^{in}$}}
\put(2457,1550){\makebox(0,0)[r]{$r_H^{out}$}}
\put(587,1423){\makebox(0,0)[r]{$\bullet$}}
\put(1258,1423){\makebox(0,0)[r]{$\bullet$}}
\put(647,1523){\makebox(0,0)[r]{$r_{top}$}}
\put(1147,1300){\makebox(0,0)[r]{$r_{top}=r_{min}$}}
\put(1347,1500){\makebox(0,0)[r]{$r_{min}$}}
\put(1047,1500){\makebox(0,0)[r]{$r_{0}$}}
\put(2847,1500){\makebox(0,0)[r]{$r_{max}$}}
\put(1058,1423){\makebox(0,0){$\bullet$}}
\put(1058,1423){\vector(0,-1){350}}
\end{picture}

\vskip 2em
\caption{Schematic shapes of the effective potential $V(r)$ for the
radial motions of the massive particle. The shaded region between the inner
and outer horizons is forbidden for the test particle. }
\label{fig8}
\end{figure}

\begin{center}{
\begin{tabular}{|c|c|c|} \hline
Fig.~8 & $\gamma$ & orbit$\;$force\\ \hline
(i) & $\gamma<e^{N(0)}$ & $r_{min}\leq r<r^{in}_{H}$(repulsive), 
$r^{out}_{H}<r\leq r_{max}$(attractive)\\ \hline
(ii) & $\gamma=e^{N(0)}$ & $r_{min}\leq r<r^{in}_{H}$(repulsive),
$r^{out}_{H}<r\leq r_{max}$(attractive)\\ 
& & Stopped motion at $r=0$\\ \hline
(iii) & $e^{N(0)}<\gamma<\gamma_{top}$ & $r_{min}\leq r<
r^{in}_{H}$(repulsive), $r^{out}_{H}<r\leq r_{max}$(attractive)\\ 
& & $r<r_{top}<r_{min}$(attractive)\\ \hline
(iv) & $\gamma=\gamma_{top}$ & $r<r_{top}$(attractive), Stopped motion at
$r_{top}$\\
& & $r<r^{in}_{H}$(repulsive), $r^{out}_{H}<r\leq r_{max}$(attractive)\\ \hline
(v) & $\gamma_{top}<\gamma<\gamma_{flat}$ &
$r<r^{in}_{H}$(attractive-repulsive), $r^{out}_{H}<r<r_{max}$(attractive)
\\ \hline
(vi) & $\gamma=\gamma_{flat}$ & $r<r^{in}_{H}$(no force at $r_0$, 
attractive elsewhere)\\ 
& & $r^{out}_{H}<r<r_{max}$(attractive)\\ \hline
(vii) & $\gamma_{flat}<\gamma$ & $r<r^{in}_{H}$, $r^{out}_{H}<r<r_{max}$
(attractive everywhere)
\\ \hline
\end{tabular}
}\end{center}

\vspace{5mm}

\noindent Table 2. The radial motions of a test particle of mass $m=1$ for
various $\gamma$'s, when $|\Lambda|/\lambda v^{2}=$1.4, $Gv^{2}=$0.1.

\vspace{5mm}

B-II(III)-(d) \underline{$(m=1,L\ne 0)$}: For any rotating motion for the
massive particle under the influence of $B(r)$ for the vortex-black hole, the
allowed regions are $r_{min}\leq r<r^{in}_{H}$ in which the force is
repulsive, and $r^{out}_{H}<r\leq r_{max}$ in which the force is attractive
(See the dashed line in Figure 7).

\vspace{10mm}

Under the metric written as in Eq.~(\ref{metr})
the conserved quasilocal mass measured by static observer at $r$
is given by~\cite{brown}
\begin{eqnarray}\label{quasi}
8G M_{q}=  2\sqrt{e^{2N(r)}B(r)}\left( \sqrt{B_0(r)} -\sqrt{B(r)}\right).
\end{eqnarray}
Here $B_0(r)$ is the background metric $g^{rr}$ which determines
the zero point of energy. 
The background can be obtained simply by setting integration constant of 
a particular solution to some specific value that specifies the
reference frame. 
As we discussed previously, the background is the spacetime without the
global vortex, specifically, $n=0$ and $|\phi|(r)=v$ and thereby
$B_{0}(r)=|\Lambda|r^{2}+1$.
When the spacetime is asymptotically flat, 
the usual Arnowitt-Deser-Misner(ADM) mass $M_q$ 
is determined in the limit $r \rightarrow \infty$.
For sufficiently large $r$, Eq.~(\ref{ninf}) and Eq.~(\ref{binf}) give 
\begin{eqnarray}
M_q\stackrel{r\rightarrow\infty}{\longrightarrow}\pi n^{2}v^{2}
\ln r/r_{c} + {\cal M}.
\end{eqnarray}
Thus the quasilocal mass determined at $r\rightarrow\infty$ 
contains two terms: finite negative mass from the core of the vortex
and the logarithmically divergent one from the 
topological sector of Goldstone degree.
This coincides approximately with the mass formula of the global vortex
in flat spacetime, which is obtained by the spatial integration of the
time-time component of energy-momentum tensor. Though $M_{q}$ cannot be 
identified as ADM mass due to the hyperbolic structure of global spacetime
of our interest, its form looks natural once we recall nonpropagation 
of (2+1) dimensional graviton in anti-de Sitter gravity.

\setcounter{section}{4}
\setcounter{equation}{0}
\begin{center}\section*{\large\bf IV. Topological Charge as a Black Hole Charge}
\end{center}
\indent\indent 
In the previous section we have shown that 
the long tail of neutral static vortex can provide
the black hole charge in (2+1)-dimensional BTZ black hole. By use of
the duality transformation \cite{KL}, we construct the direct 
relationship between the topological charge $n$ of the neutral vortex 
and the electric charge of the dual transformed theory. Through this
analysis the reason why the vorticity $n$ of neutral objects can play the
same role of (electric) charge of the BTZ black hole will be manifested.

The path integral of our theory is written as
\begin{eqnarray} \label{Dual}
Z&=&\int [dg_{\mu\nu}][d\bar{\phi}][d\phi]\exp i\biggl\{\int d^{3}x\!
\sqrt{g}\Bigl[ -\frac{1}{16\pi G}(R+2\Lambda )+\frac{1}{2}g^{\mu\nu}
\partial_{\mu}\bar{\phi}\partial_{\nu}\phi-V(|\phi|)\Bigr]\biggr\}.
\end{eqnarray}
Rewrite the scalar field in the path integral in terms of radial variables
$\phi=|\phi|e^{i\Omega}$
and linearize the term of scalar phase such as
\begin{eqnarray}
\lefteqn{\int[d\Omega]\exp\biggl\{ i\int d^{3}x\sqrt{g}
\frac{g^{\mu\nu}}{2} |\phi|^{2}\partial_{\mu}\Omega
\partial_{\nu}\Omega\biggr\} }\nonumber\\
&=&\prod_{x}|\phi|^{-3}g^{\frac{1}{4}}
\int[d\Omega][dC_{\mu}]\exp\biggl\{ i\int d^{3}x\sqrt{g}
\Bigl[-\frac{g^{\mu\nu}}{2}\Big(\frac{C_{\mu}C_{\nu}}{|\phi|^2}
-2C_{\mu}\partial_{\nu}\Omega\Big)\Bigr]\biggr\}.
\label{pathint2}
\end{eqnarray}
Let us divide the configurations of the 
scalar phase by the topological sector
$\Theta$ which is $\frac{\epsilon^{\mu\nu\rho}}{\sqrt{g}}\partial_{\nu}
\partial_{\rho}\Theta\ne 0$, and the single-valued part $\eta$ which satisfies
$\frac{\epsilon^{\mu\nu\rho}}{\sqrt{g}}\partial_{\nu}\partial_{\rho}\eta=0$:
$\Omega=\Theta+\eta$ and $[d\Omega|=[d\Theta][d\eta]$.
Integrating out $\eta$ and using $[d\partial_{\mu}\eta]=[d\eta]$ up to a
field-independent Jacobian factor, we have 
\begin{eqnarray}\label{del}
(\ref{pathint2})&=&
\prod_{x}|\phi|^{-3}g^{\frac{1}{4}}\int[d\Theta][d\eta_{\mu}][dC_{\mu}]
\,\delta(\frac{\epsilon^{\mu\nu\rho}}{\sqrt{g}}\partial_{\nu}\eta_{\rho})
\nonumber\\
&&\times\exp\biggl\{i\int d^{3}x\sqrt{g}
\Bigl[-\frac{g^{\mu\nu}}{2|\phi|^{2}}C_{\mu}
C_{\nu}+g^{\mu\nu}C_{\mu}(\partial_{\nu}\Theta+\eta_{\nu})\Bigr]\biggr\}.
\end{eqnarray}
Let us rewrite the delta functional in Eq.~(\ref{del}) by introducing the dual
vector field $A_{\mu}$, i.e., $\delta(\frac{\epsilon^{\mu\nu\rho}}{\sqrt{g}}
\partial_{\nu}\eta_{\rho})={\displaystyle\int}
[dA_{\mu}]\exp\bigl\{-iv\int d^{3}x\sqrt{g}
~\frac{\epsilon^{\mu\nu\rho}}{\sqrt{g}}A_{\mu}\partial_{\nu}\eta_{\rho}\bigr\}$,
and integrate out $\eta_{\mu}$. Then we obtain a relation from the delta
functional, $C_{\mu}=\frac{v}{2}\sqrt{g}\epsilon_{\mu\nu\rho}F^{\nu\rho}$ and
$F_{\mu\nu}=\partial_{\mu}A_{\nu}-\partial_{\nu}A_{\mu}$. Finally if we do the 
integration over the vector auxiliary field $C_{\mu}$, then Eq.~(\ref{Dual}) 
becomes
\begin{eqnarray} \label{Max}
Z&=&\int [g^{\frac{3}{4}}dg_{\mu\nu}][|\phi|^{-2}d|\phi|][dA_{\mu}][d\Theta]
\exp\biggl\{i\int d^{3}x\sqrt{g}\Bigl[-\frac{1}{16\pi G}(R+2\Lambda)\nonumber\\
&&\hspace{10mm}+\frac{1}{2}g^{\mu\nu}\partial_{\mu}|\phi|\partial_{\nu}|\phi|
-V(|\phi|)
-\frac{v^{2}}{4|\phi|^{2}}g^{\mu\nu}g^{\rho\sigma}F_{\mu\rho}F_{\nu\sigma}
+\frac{v\epsilon^{\mu\nu\rho}}{2\sqrt{g}}F_{\mu\nu}\partial_{\rho}\Theta\Bigr]
\biggr\}.
\end{eqnarray} 
This duality transformation can be achieved in arbitrary (D$+$1)
dimensions by use of antisymmetric tensor field of rank (D$-$1),
so the Maxwell-like term in Eq.~(\ref{Max}) becomes nothing but the
Kalb-Ramond action~\cite{Wit} in (2$+$1)D.
Euler-Lagrange equations read
\begin{eqnarray} 
\frac{1}{\sqrt{g}}\partial_{\mu}(\sqrt{g}g^{\mu\nu}\partial_{\nu}|\phi|)
=\frac{v^{2}}{2|\phi|^{3}}g^{\mu\nu}g^{\rho\sigma}F_{\mu\rho}F_{\nu\sigma}
-\frac{dV}{d|\phi|}
\end{eqnarray}
\begin{eqnarray}\label{gaugeeq}
\frac{1}{\sqrt{g}}\partial_{\nu}\Bigl(\sqrt{g}\frac{v^{2}}{|\phi|^{2}}
F^{\mu\nu}\Bigr)=
\frac{\epsilon^{\mu\nu\rho}}{\sqrt{g}}\partial_{\nu}\partial_{\rho}
\Theta
\end{eqnarray}
\begin{eqnarray}\label{deineq}
\lefteqn{R_{\mu\nu}-\frac{g_{\mu\nu}}{2}(R+2\Lambda)}\nonumber\\
&=&\frac{v^{2}}{4|\phi|^{2}}g^{\rho\sigma}(g_{\mu\nu}g^{\tau\kappa}
-4g^{\tau}_{\;\mu}g^{\kappa}_{\;\nu})F_{\rho\tau}F_{\sigma\kappa}
+(g^{\rho}_{\;\mu}g^{\sigma}_{\;\nu}-\frac{1}{2}g_{\mu\nu}g^{\rho\sigma})
\partial_{\rho}|\phi|\partial_{\sigma}|\phi|+g_{\mu\nu}V.
\end{eqnarray}

Since we are interested in the 
neutral objects which do not carry global $U(1)$ charge
($C_{0}=g_{0\mu}|\phi|^{2}\partial^{\mu}\Omega=0$), they do not
carry dual magnetic field ($F^{ij}=\frac{\epsilon^{0ij}}{\sqrt{g}}C_{0}=0$).
Thus the spatial components of the equation for the dual gauge field
are automatically satisfied. The time component of Eq.~(\ref{gaugeeq}) is
nothing but the Gauss' law in asymptotic region for large
$r$ ($|\phi|\rightarrow v$): For the rotationally symmetric 
vortex solutions $\Theta=n\theta$, it is 
\begin{eqnarray}\label{gausseq}
\frac{1}{\sqrt{g}}\partial_{i}\Bigl(\sqrt{g}F^{0i})
\approx n\frac{1}{\sqrt{g}}\delta^{(2)}(\vec{x}).
\end{eqnarray}
The next order term of the scalar amplitude due to the small perturbation 
from the vacuum value $v$ does not contribute to the charge, so one can easily 
identify the vorticity $n$ as the electric charge of the dual gauge field.
Similarly, since the scalar amplitude terms, 
which are the second and third terms 
in the right-hand side of Eq~(\ref{deineq}), fall 
rapidly as the radial coordinate $r$ increases,
time-time component of Einstein equations in Eq.~(\ref{deineq}) has the leading
contribution from the negative cosmological constant term and the next 
leading term from the electric energy for large $r$:
\begin{eqnarray}
G_{00}\approx e^{2N}B\Big(\Lambda+\frac{v^{2}}{2|\phi|^{2}}(F_{0i})^{2}\Big)
\sim e^{2N}B\Big(\Lambda+v^{2}\frac{n^{2}}{r^{2}}\Big).
\label{G00}
\end{eqnarray}
Obviously, the electric field can be identified as that of the point charge at
the origin. The self-energy in flat spacetime contains logarithmic divergence.
Therefore this topological charge can constitute the charge of the
BTZ black holes: 
$B(r)\approx |\Lambda|r^2-8\pi G v^2 n^2 \ln r -{\cal M}$ for large $r$.

At the core of the vortex, the nonvanishing component of the dual electric
field $F_{0r}$ is regular: For small $r$,
\begin{eqnarray}
F_{0r}\sim -n\frac{\phi_0}{v^2} r^{2n-1}, 
\end{eqnarray}
since $|\phi|(r)\sim \phi_0 r^n$. The dual electric field term of the
energy-momentum tensor in Eq.~(\ref{G00}) is also regular;
\begin{eqnarray}
T_{00}&\propto& e^{2N}B\frac{v^2}{2|\phi|^2} (F_{0r})^2     \nonumber \\
      &\sim& \frac{n^2}{2}\frac{\phi_0^2}{v^2} e^{2N(0)} r^{2(n-1)}.
\end{eqnarray}
Therefore, the role of $1/|\phi|^{2}$ in Eq.~(\ref{gaugeeq}) and 
Eq.~(\ref{deineq}) is a regulator of the soliton at its core.

Now we have an understanding that the addition of the global vortex of
vorticity $n$ to the center of the Schwarzschild-type BTZ black hole produces
a Reissner-Nordstr\"om-type BTZ black hole of electric charge $n$. 
This implies
that spinless static vortices with finite energy in flat spacetime, {\it e.g.},
the topological charge of Abrikosov-Nielsen-Olesen vortices in Abelian Higgs 
model or that of topological  lumps in $O(3)$ nonlinear sigma model 
can not give rise to an additional BTZ black hole (electric) charge
since they do not carry long tail of energy density. 
However, it is an open question that whether the spinning charged solitons, 
{\it e.g.} $Q$-lumps (or nontopological global vortices)~\cite{Leese} 
or topological or nontopological vortices in Chern-Simons theories
\cite{HKP,CCK}, can constitute an BTZ black hole with both charge and spin. 

\setcounter{section}{5}
\setcounter{equation}{0}
\begin{center}\section*{\large\bf V. Physical Relevance as a Black Cosmic 
String in (3+1)D}
\end{center}
\indent\indent 
We have considered  the global vortices in (2+1)D 
curved spacetime, however
these point particle like extended objects on spatial plane may describe the
straight global $U(1)$ strings along the $z$-direction \cite{VS}.
If we consider a static metric of cylindrically symmetric 
string along the $z$-axis 
\begin{eqnarray}
ds^{2}=B(r)e^{2N(r)}(dt^{2}-dz^{2})-\frac{dr^{2}}{B(r)}-r^{2}d\theta^{2}
\end{eqnarray}
which also has boost invariance in $z$-direction,
the previous analysis moves to (3+1) dimensional anti-de Sitter spacetime with
no change since we already adjusted the dimension of fields and constants to
those in (3+1) dimensions. 

Here let us take into account a perfect  situation: A
global $U(1)$ static string straight along $z$-axis was generated in some
symmetry breaking scale $v$ and has evolved safely to a static object in the
present universe. Inserting the Newton constant and the present lower bound of 
cosmological constant into Eq.~(\ref{b22}), we have the critical value for 
extremal black hole in our room temperature scale, 
${|\Lambda|}/{2\pi\lambda v^{4}}\approx 0.3{\rm eV}$. 
This implies that, when the
cosmological constant is negative and bounded by the experimental lower limit
in the present universe 
($-(0.34\sim0.99)\times10^{-83}({\rm GeV})^2<\Lambda<(0.68\sim1.98)\times
10^{-83})({\rm GeV})^2$~\cite{Dat}), the global strings produced 
almost all the scales
remain as charged black strings. However, the global vortices made in 
``relativistic" ${}^{4}$He superfluid are regular~\cite{Vol}. 
The characteristic scale $r_{H}$ in 
Eq.~(\ref{rhor}) is $10^{6}$ pc for grand unified scale $v\sim 10^{15}$GeV, 
and is $10^{-2}$ A.U. for electroweak scale. The underlying physics for the
reason why we reached this enormous size of horizon is easy: The mass density
of black cosmic string per unit length is given by the ratio of scalar mass
and the Planck scale, $2\sqrt{\pi G}v$, but the negative vacuum energy density
inside the horizon is given by the ratio of the square root of the absolute
value of cosmological constant and the scalar mass,
$\sqrt{|\Lambda|}/\sqrt{\lambda}v$. The scale of the cosmic string generation
is large, but the lower bound of present cosmic vacuum energy is extremely
small. Then the scale for this black cosmic string characterized by
the horizon scale should be very large. 
Though these values are obtained under a perfect presumed toy situation
without taking into account fluctuations around the black cosmic string, the huge
radius of it, namely, the radius of the black cosmic string produced in GUT scale ($\sim 10^6$pc)
is larger than the diameter of our galaxy ($\sim 5\times 10^4$pc), may
imply difficulty for the survival of the charged black cosmic strings produced
in such early universe in the present universe with extremely small
bound for the cosmological constant.
Once a global cosmic string is produced, it starts to radiate gapless
Goldstone bosons. This dominant mechanism for energy loss makes the life time
of a typical string loop very short \cite{Dav}: A global string loop
oscillates about 20 times before radiating most of its energy which is
contrasted with gravitational radiation where the oscillation lasts about 
$10^{4}$ times.
The space outside the horizon of black cosmic string 
is almost flat except for tiny attractive force due to negative cosmological 
constant as shown in Eq.~(\ref{radial}) and Eq.~(\ref{binf}), 
and then the massless Goldstone bosons
can be radiated outside the horizon. 
However, almost all the energy accumulated inside 
the horizon remains eternally.
This ``black" nature of the global $U(1)$ string 
in the anti-de Sitter spacetime is remarkable at least for the case of 
straight cosmic strings.

{}Finally, let us emphasize again that we have two mass scales, the core mass
and the inverse of the horizon, which are determined by three energy scales of
big difference, namely the Plank scale ($1/\sqrt{G}\sim 10^{19}$GeV),
the present bound of the cosmological constant 
($\sqrt{|\Lambda|}\sim 10^{-42}$GeV), and the symmetry breaking scale
 (from $v\sim 10^{19}$GeV to $v\sim 0.3$eV).
Therefore, the very existence of this horizon is expected to 
change drastically the physics related to the dynamics of global $U(1)$ 
strings, {\it e.g.}, the intercommuting of two strings or 
the production of wakes by moving long strings \cite{VS}.

\setcounter{section}{6}
\setcounter{equation}{0}
\begin{center}\section*{\large\bf VI. Conclusion}
\end{center}
\indent\indent In this paper, we have considered a scalar field model with a
spontaneously broken $U(1)$ global symmetry in (2+1) dimensional anti-de
Sitter spacetime, and investigated the cylindrically symmetric vortex
solutions. We have found regular topological soliton configurations of which
base manifolds constitute smooth hyperbolic space, extremal BTZ black hole and
charged BTZ black hole according to the decreasing magnitude of negative
cosmological constant. Different from the zero cosmological constant space 
supported by the global $U(1)$ vortex, which cannot avoid the physical
singularity, the obtained anit-de Sitter spaces are (physical) singularity
free. Due to the logarithmic long tail of the Goldstone mode, the BTZ black
hole also carries the charge, which is identical to the case of an electric
Maxwell charge. This identification was constructed by the duality
transformation. 
All possible geodesic motions of massive and massless test particles were 
analyzed. Since the asymptotic space is hyperbolic, all the motions of massive
particles  are bounded. However, some massless test particles can escape
to the spatial infinity of hyperbola.

In (3+1) dimensions, the obtained global vortex-BTZ black hole depicts a 
straight charged black cosmic string. We brought up a toy model situation
that these objects formed through a cosmological phase transition in the early
universe (from the grand unification scale to the standard model scale) and
survive in the present universe assumed with allowably small magnitude of the
negative cosmological constant ($|\Lambda|\sim 10^{-83}$GeV${}^{2}$). The
corresponding scale of horizon $r_{H}$ is in order from $10^{6}$pc to 
$10^{-2}$A.U.. Then it implies that the observation of black cosmic string in the
present universe may relate the bound of negative cosmological constant to
the production of global $U(1)$ vortices in the early universe.

Three brief comments are now in order. (i)
For the vortices in Abelian Higgs model or $O(3)$ nonlinear sigma model, they
have finite energy in flat spacetime. 
A question of interests is whether they can form the black
holes in anti-de Sitter space. 
Until now we do not have an answer to this question~\cite{KM}.
If we find them, such BTZ black holes must be Schwarzschild type 
without electric charge. 
(ii) Static charged BTZ black holes can also be obtained in dilaton gravity.
Therefore, the global $U(1)$ vortices coupled to dilaton and
anti-de Sitter gravity may have some relevance in stringy cosmology \cite{CM}.
(iii) For more realistic models of straight static black cosmic strings,
the general metric of the form
$ ds^2 =B(r) e^{2N(r)}(dt-C(r)dz)^2-B(r)dr^2 -r^2d\theta^2 -D(r)dz^2$
has to be taken into account.
\vspace{5mm}

\begin{center}{\large\bf Acknowledgments}
\end{center}

This work was 
supported by the Ministry of Education(BSRI/97-2418) and 
the KOSEF(95-0702-04-01-3 and through CTP, SNU).

\def\hebibliography#1{\begin{center}\subsection*{References
}\end{center}\list
  {[\arabic{enumi}]}{\settowidth\labelwidth{[#1]}
\leftmargin\labelwidth	  \advance\leftmargin\labelsep
    \usecounter{enumi}}
    \def\newblock{\hskip .11em plus .33em minus .07em}
    \sloppy\clubpenalty4000\widowpenalty4000
    \sfcode`\.=1000\relax}

\let\endhebibliography=\endlist

\begin{hebibliography}{100}
\bibitem{DJH} A. Staruszkiewicz, Acta. Phys. Polo. {\bf 24}, 735 (1963); 
S. Deser, R. Jackiw and G. 't Hooft, ANN. Phys. {\bf 152}, 220 (1984); J. Gott 
and M. Alpert, Gen. Rel. Grav. {\bf 16}, 243 (1984); S. Giddings, J. Abbot and
K. Kuchar, Gen. Rel. Grav. {\bf 16}, 751 (1984).
\bibitem{DJ} S. Deser and R. Jackiw, Ann. Phys. {\bf 153}, 405 (1984).
\bibitem{BTZ} M. Ba\~{n}ados, C. Teitelboim and J. Zanelli, Phys. Rev. Lett.
{\bf 69}, 1849 
(1992); M. Ba\~{n}ados, M. Henneaux, C. Teitelboim and J. Zanelli, 
             Phys. Rev. D {\bf 48}, 1506 (1993).
\bibitem{VS} For a review, see A. Vilenkin and E.P.S. Shellard, {\it Cosmic
Strings and Other Topological Defects}, (Cambridge, 1994); 
M.B. Hindmarsh and T.W.B. Kibble, Rept. Prog. Phys. {\bf 58}, 477 (1995).
\bibitem{LNW} K. Lee, V.P. Nair and E. Weinberg, Phys. Rev. Lett. {\bf 68},
1100 (1992); Phys. Rev. D {\bf 45}, 2751 (1992); M.E. Ortiz, Phys. Rev. D {\bf
45}, R2586 (1992); P. Breitenlohner, P. Forg\'{a}cs, and D. Maison,
Nucl. Phys. B {\bf 383}, 357 (1992);
For a review, see E. Weinberg, in Proceedings of the
13th Symposium on Theoretical Physics 
edited by J.E. Kim, (Min Eum Sa, Seoul, 1995), and 
K. Maeda, in Proceedings of the First
Seoul Workshop on Gravity and Cosmology edited by S.W. Kim, P. Oh and J. Lee,
(Korean Physical Society, Seoul, 1995).
\bibitem{HS} 
A. Vilenkin and A.E. Everett, Phys. Rev. Lett. {\bf 48}, 1867 (1982);
E.P.S. Shellard, Nucl. Phys. B {\bf 283}, 624 (1987); 
D. Harari and P. Sikivie, Phys. Rev. D {\bf 37}, 3438 (1988).
\bibitem{Gre} R. Gregory, Phys. Lett. B {\bf 215}, 663 (1988);
Cohen and D.B. Kaplan, Phys. Lett. B {\bf 215}, 67 (1988);
G.W. Gibbons, M.E. Ortiz and F. Ruiz Ruiz, Phys. Rev. D {\bf 39}, 1546 (1989).
\bibitem{KKK} N. Kim, Y. Kim and K. Kimm, Preprint
                     SNUTP/97-022, gr-qc/9707011.
\bibitem{KK} If we solve the static objects with vanishing 
$i0$- and $ij$-components of energy-momentum tensor, {\it e.g.}, massive but spinless
point particles or self-dual vortices without angular momentum, in 
anti-de Sitter spacetime, the stationary metric solutions contain
another  possibility in addition to two known solutions
(Deser-Jackiw type or BTZ type). It will be discussed in a paper by
Y. Kim and K. Kimm, in preparation.
\bibitem{Inc} E.L. Ince, see chapter XIV (page 334) in 
   {\it Ordinary Differential Equations}, (Dover, New York, 1956).
\bibitem{Cle} G. Clement, Phys. Rev. D {\bf 50}, 7119 (1994);
                          ~Phys. Lett. B {\bf 367}, 70 (1996);
J.S.F. Chan, K.C.K. Chan and R.B.  Mann, Phys. Rev. D {\bf 54}, 1535 (1996).
\bibitem{HL} D. Harari and C. Loust\'{o}, Phys. Rev. D {\bf 42}, 2626 (1990);
             Y. Kim, K, Maeda and N. Sakai, Nucl. Phys. B {\bf 481},
              453 (1996).
\bibitem{CMP} 
C. Farina, J. Gamboa and A.J. Segui-Santonja, 
        Class. Quant. Grav. {\bf 10}, L193 (1993);
N. Cruz, C. Mart\'inez and L. Pe\~na, Class. Quant. Grav. {\bf 
11}, 2731 (1994).
\bibitem{brown} J.D. Brown and J.W. York, Phys. Rev. D {\bf 47}, 1407 (1993);
                J.D. Brown, J. Creighton, and R.B. Mann, 
                Phys. Rev. D {\bf 50}, 6394 (1994).
\bibitem{KL} Y. Kim and K. Lee, Phys. Rev. D {\bf 49}, 2041 (1994); 
K. Lee, Phys. Rev. D {\bf 49}, 4265 (1994); C. Kim and Y. Kim, Phys. Rev. D 
{\bf 50}, 1040 (1994).
\bibitem{Wit} E. Witten, Phys. Lett B {\bf 153}, 243 (1985);
R. L. Davis and E.R.S. Shellard, Phys. Lett. B {\bf 214}, 219 (1988).
\bibitem{Leese} R.A. Leese, Nucl. Phys. B {\bf 366}, 283 (1991);
 C. Kim, S. Kim and Y. Kim, Phys. Rev. D {\bf 47}, 5434 (1993).
\bibitem{HKP} J. Hong, Y. Kim and P.Y. Pac, Phys. Rev. Lett. {\bf 64}, 
              2230~(1990); R. Jackiw and E.J. Weinberg, {\it ibid}, {\bf 64},
              2234~(1990); R. Jackiw, K. Lee and E.J. Weinberg,
              Phys. Rev. D {\bf 42}, 3488~(1990).
\bibitem{CCK} P. Valtancoli, Int. J. Mod. Phys. A {\bf 18}, 4335 (1992); D.
Cangemi and C. Lee, Phys. Rev. D {\bf 46}, 4768 (1992); G. Cl\'{e}ment, Phys.
Rev. D {\bf 54}, 1844 (1996).
\bibitem{Dat} Phys. Rev. D {\bf 54}, 66 (1996) Part I, Review of Particle 
Physics.
\bibitem{Vol} Suppose that the model of our consideration is relativistic 
counterpart of superfluid models and then we can use the phase transition
scales for $^4{\rm He} (\sim 10^{-4}$ eV) and $^3 {\rm He} (\sim 10^{-7}$ eV) 
as values
for $v$. From the critical condition for black vortex formation 
$(| \Lambda | \approx 2\pi G \lambda v^4$ from Eq.(\ref{rhor})), we can compute the 
bounds of the 
negative cosmological constant : For $\lambda=1$, $|\Lambda | > 10^{-110}
({\rm GeV})^2$ for $^4 {\rm He}$ and $| \Lambda | > 10^{-122} ({\rm GeV})^2$
for ${}^{3}$He in order not to form the black vortex.
Of course, this naive result should be refined by taking into account
realistic condensed matter models, specific form of gravitational interaction, and
systematic nonrelativistic expansion. However, our study may have some 
relevance to the works by G.E. Volovik (See preprints cond-mat/9706149, 
cond-mat/9706172).
\bibitem{Dav} R.L. Davis, Phys. Rev. D {\bf 32}, 3172 (1985); A. Vilenkin and
T. Vachaspati, Phys. Rev. D {\bf 35}, 1138 (1987); For a review, see
Ref.~\cite{VS} and the references therein.
\bibitem{KM} Y. Kim and S. Moon, in preparation.
\bibitem{CM} N. Kaloper, Phys. Rev. D {\bf 48}, 2598 (1993);
K.C.K. Chan and R.B. Mann, Phys. Rev. D {\bf 50}, 6385 (1994),
Erratum-{\it ibid}~D {\bf 52}, 2600. 
\end{hebibliography}

\end{document}